\documentclass[aps,prd,preprint,superscriptaddress,amsmath,amssymb,showpacs]{revtex4-1}
\usepackage{dcolumn}
\usepackage{graphicx}
\usepackage{float}
\usepackage{physics}
\usepackage[colorlinks=true,allcolors=blue]{hyperref}
\usepackage{mathrsfs}
\usepackage{inputenc}
\usepackage{diagbox}

\begin{document}
\title{Inverse magnetic catalysis and energy loss in a holographic QCD model}

\author{Zhou-Run Zhu}
\email{zhuzhourun@mails.ccnu.edu.cn}
\affiliation{Institute of Particle Physics and Key Laboratory of Quark and Lepton Physics (MOS), Central China Normal University,
Wuhan 430079, China}

\author{Defu Hou }
\thanks{Corresponding author}
\email{houdf@mail.ccnu.edu.cn}
\affiliation{Institute of Particle Physics and Key Laboratory of Quark and Lepton Physics (MOS), Central China Normal University,
Wuhan 430079, China}

\date{\today}

\begin{abstract}
In this paper, we consider the Einstein-Maxwell-dilaton holographic model for light quarks with nonzero magnetic field and chemical potential. First, we study the phase diagrams in $T-\mu$ and $T-B$ planes. We observe inverse magnetic catalysis which is consistent with the lattice QCD results. We discuss the influence of the magnetic field and chemical potential on the location of the critical end point (CEP). It is found that the magnetic field increases the critical $\mu_{\scriptscriptstyle CEP}$ of the CEP in the $T-\mu$ plane and the chemical potential increases the critical $B_{\scriptscriptstyle CEP}$ of the CEP in the $T-B$ plane. Second, we discuss the equations of state (EOS) with nonzero magnetic field and chemical potential. We observe that the EOS near the phase transition temperature are nonmonotonic. Then we study the energy loss with a nonzero magnetic field and chemical potential. It is found that the drag force of the heavy quark and jet quenching parameter $\hat{q}$ show an enhancement near the phase transition temperature. The peak values of drag force and $\hat{q}$ are pushed toward lower temperature with increasing $B$ or $\mu$. This phenomenon is consistent with the phase transition temperature decrease with increasing $B$ or $\mu$ in this holographic model. Moreover, we find that the heavy quark may lose more energy when it is perpendicular to a magnetic field which is consistent with the results of the jet quenching parameter.
\end{abstract}

\maketitle

\section{Introduction}\label{sec:01_intro}
The experimental investigation of the phase structure of quantum chromodynamics (QCD) matter is one of the important tasks of heavy ion collision experiments at the Relativistic Heavy Ion Collider (RHIC) and LHC \cite{Arsene:2004fa,Adcox:2004mh,Back:2004je,Adams:2005dq,Aad:2013xma}. It is generally believed that the strongly coupled plasma can be generated from experiments since heavy ion collisions create extreme conditions of high temperature and energy density. It is well known that QCD matter is in the confinement phase in the region of low temperature $T$ and small chemical potential $\mu$ while it is in the deconfinement phase in the high temperature $T$ and large chemical potential $\mu$ region. Furthermore, the phase diagram for light quarks \cite{Philipsen:2010gj} cross over at $\mu < \mu_c$ (critical chemical potential) and becomes first-order at $\mu$ $>$ $\mu_c$. It is interesting and challenging to probe the phase diagram of QCD. The phase diagram in the $T- \mu$ plane can provide  rich information for promoting the understanding of the properties of QCD matter under extreme conditions\cite{Stephanov:2004wx,Fukushima:2013rx}.

The phase structure of QCD under an external magnetic field has attracted much attention since a strong magnetic field has been generated in the early stage of noncentral heavy-ion collisions \cite{Kharzeev:2007jp,Skokov:2009qp,Deng:2012pc} and the early Universe \cite{Vachaspati:1991nm}. Although the magnetic field rapidly decays, it is still intense at the initial formation of the quark-gluon plasma (QGP) \cite{Tuchin:2013apa,McLerran:2013hla}. As expected, the strong magnetic field has affected the plasma evolution and the charge dynamics in the strongly interacting matter. Studying QCD properties in the magnetized system is still a major focus of current research, especially the investigation of the phase structure \cite{DElia:2010abb,Kharzeev:2012ph,Miransky:2015ava}. Lattice QCD serves as a powerful tool to explore the QCD phase structure under magnetic field. Early lattice QCD results indicated the chiral condensate and phase transition temperature increase with the magnetic field, namely magnetic catalysis (MC) \cite{Shovkovy:2012zn,Ilgenfritz:2012fw}. The opposite behaviors were observed in further lattice QCD studies with physical light quark masses, so-called inverse magnetic catalysis (IMC) \cite{Bali:2011qj,Bali:2012zg,Ilgenfritz:2013ara}. Using the improved discretization schemes, IMC has been found in the lattice QCD investigations \cite{Bornyakov:2013eya,Bali:2014kia,Tomiya:2019nym}. Further intensive study of (inverse) magnetic catalysis has drawn much attention to unveil the microscopic mechanism of the magnetic field effects \cite{DElia:2011koc,Bruckmann:2013oba,Fukushima:2012kc,Fraga:2012fs,Bali:2013esa,Ayala:2014iba,Andersen:2014xxa,Ferreira:2014kpa,Yu:2014sla,Feng:2015qpi,Ding:2020hxw}.
It is found that the (inverse) magnetic catalysis is related to the quark mass \cite{DElia:2018xwo,Endrodi:2019zrl}. Inverse magnetic catalysis is valid for light quarks. The system turns into MC from IMC with the increase of the quark mass.

It is expected that IMC is mainly caused by the strongly coupled dynamics around phase transition. The perturbative QCD method is not reliable since the coupling constant is still large around the phase transition temperature. Furthermore, lattice QCD does not work well at finite chemical potential due to the sign problem. Studying the QCD phase structure from AdS/CFT correspondence \cite{Witten:1998qj,Gubser:1998bc,Maldacena:1997re} may provide some significant inspiration. The characteristic features of QCD and phase structure at finite temperature and chemical potential have been discussed by using this powerful nonperturbative approach \cite{DeWolfe:2010he,DeWolfe:2011ts,Cai:2012xh,Finazzo:2016psx,Knaute:2017opk,Sin:2007ze,Colangelo:2010pe,Ballon-Bayona:2020xls,He:2013qq,Yang:2015aia,Yang:2014bqa,Li:2017tdz}.
The confinement-deconfinement phase transition and QCD thermodynamic properties under an external magnetic field also have been widely investigated from holography and some interesting results have been obtained \cite{Johnson:2008vna,DHoker:2009ixq,Preis:2010cq,Alam:2012fw,Callebaut:2013ria,Mamo:2015dea,Rougemont:2015oea,Chelabi:2015cwn,Dudal:2015wfn,Li:2016gfn,Rodrigues:2017iqi,Gursoy:2016ofp,Gursoy:2017wzz,Ballon-Bayona:2022uyy}. In this work, we want to study the influence of the magnetic field
and chemical potential on the location of critical end point (CEP) simultaneously. To be specific, we want to discuss the effects of the magnetic field on the critical $\mu_{\scriptscriptstyle cep}$ in the $T-\mu$ plane and the chemical potential on the critical $B_{\scriptscriptstyle CEP}$ in the $T-B$ plane. It is worth mentioning that the bottom-up holographic QCD models can describe QCD-like physics successfully \cite{Li:2011hp,Cai:2012eh,Dudal:2017max,Bohra:2019ebj,Bohra:2020qom,He:2020fdi,Arefeva:2020byn,Arefeva:2022avn}.
The (inverse) magnetic catalysis can be observed from the holographic QCD model. The dilaton field in the Einstein-Hilbert action is dual to the running of the coupling constant and the expression of the dilaton field is correctly solved by the gravity equations. The dynamical breaking of conformal symmetry is realized by the nontrivial profile of the dilaton field in bottom-up holographic QCD models. In bottom-up holographic QCD, one can fit the model parameters to match the properties of real QCD such as the deconfinement phase transition, equations of state, and Regge trajectory of meson mass spectrum.

When an energetic parton with a large transverse momentum passes through the QGP, it radiates gluons thereby losing energy since it interacts with the hot dense matter. This process leads to transverse momentum broadening described by the jet quenching parameter $\hat{q}$ \cite{Wang:1991xy,Baier:1996kr,Baier:1996sk,Guo:2000nz,Gyulassy:2000er,JET:2013cls,Qin:2015srf}. The jet quenching parameter plays a significant role in medium-induced radiated energy loss of light
partons and is defined by the mean squared transverse momentum per unit distance propagated, $\hat{q} = \frac{<\kappa^{2}_{\perp}>} {L}$. The holographic jet quenching parameter $\hat{q}$ was first studied in leading order of the large 't Hooft coupling \cite{Liu:2006ug} and later was extended to the subleading order in the large 't Hooft coupling at a nonzero temperature in \cite{Zhang:2012jd}. The results are in good agreement with the RHIC data \cite{Edelstein:2008cp,Zhang:2012jd} when 't Hooft coupling $\lambda = 6\pi$. The effect of the magnetic field on the jet quenching parameter from holography has been studied in \cite{Li:2016bbh,Zhang:2018pyr,Rougemont:2020had}. The temperature dependent jet quenching parameter in the pure gluon system can be seen in \cite{Li:2014dsa,Li:2014hja}. The heavy quark energy loss can be determined by drag force \cite{Gubser:2006bz,Herzog:2006gh}. In the trailing string model, drag force behaves as an observable quantity that is sensitive to the in-medium energy loss. It is the averaged momentum per unit time ($dp/dt$) and can be holographically calculated by the energy flow ($dE/dx$) from the falling string end point into the worldsheet horizon. The effect of
magnetic field or chemical potential on drag force in the EM(d) models can be seen in \cite{Rougemont:2015wca,Finazzo:2016mhm,Zhang:2018mqt,Grefa:2022sav,Zhu:2021nbl}.

In this work, we study the holographic QCD model with nonzero magnetic field and chemical potential in the Einstein-Maxwell-dilaton(EMD) gravity background. We study the phase diagram for light quarks in the $T-\mu$ and $T-B$ planes. Then we discuss the equations of state (EOS) near the phase transition temperature. In this model, we find that the magnetic field suppresses the free energy and this suppression is stronger when the connecting line of the $Q\bar{Q}$ pair is parallel to the magnetic field compared with the perpendicular case, which is consistent with the lattice QCD results \cite{Bonati:2016kxj}. The fast moving probe energy loss in the strongly coupled plasma has attracted much interest, especially near the phase transition temperature. It may be more comprehensive to consider the influence of both the magnetic field and chemical potential on energy loss. Based on this, we discuss the effects of magnetic field and chemical potential on the drag force and jet quenching parameter near the phase transition temperature simultaneously and want to characterize the phase transition temperature by energy loss.

The organization of this work is as follows. In Sec.~\ref{sec:02}, we introduce the background geometry of the AdS/QCD model with nonzero chemical potential and magnetic field. In Sec.~\ref{sec:03}, we investigate the thermodynamics in the holographic model. In Sec.~\ref{sec:04}, we discuss the free energy of the $Q\bar{Q}$ pair in the holographic model. In Sec.~\ref{sec:05}, we discuss the nonmonotonic jet quenching parameter and drag force in the AdS/QCD model. The conclusion and discussion are given in Sec.~\ref{sec:06}.

\section{Background geometry}\label{sec:02}

In this section, we first review the main derivation of the holographic AdS/QCD model with the nonzero chemical potential and magnetic field presented in \cite{Bohra:2019ebj}. The action of the Einstein-Maxwell-dilaton gravity background is
\begin{equation}
\begin{split}
\label{eqa}
 \ S = -\frac{1}{16\pi G_5 }\int d^5 x \sqrt{- g}[R-\frac{f_{1}(\phi)}{4}F_{(1)MN}F^{MN}-\frac{f_{2}(\phi)}{4}F_{(2)MN}F^{MN}-\frac{1}{2}\partial_M \phi \partial^M \phi-V(\phi) ],
 \end{split}
\end{equation}
where $\phi$ denotes the dilaton field. $V(\phi)$ represents the the potential of $\phi$. Moreover, $F_{(1)MN}$ and $F_{(2)MN}$ denote the field strength tensors of two U(1) gauge fields respectively. One can consider that the first gauge field is dual to a (neutral) flavor current which is able to create a meson. The second gauge field is dual to the electromagnetic current which can generate a different neutral meson. In \cite{Bohra:2019ebj}, the authors work with $U(1)\times U(1)$ and the meson is charge neutral. Therefore, one cannot couple electromagnetism to a meson in a direct way. In fact, there is no direct coupling between the two $U(1)$. The baryon chemical potential serves as the boundary value of time component of the first Abelian gauge field $A_{(1)M}=A_t (z)\delta^t_M$. The second Abelian gauge field can be treated as the dual of the electromagnetic current. The gauge kinetic functions $f_{1}(\phi)$ and $f_{2}(\phi)$ represent the coupling between U(1) gauge fields and the dilaton field.

The $Ans\ddot{a}tze$ of the metric \cite{Bohra:2019ebj} is
\begin{equation}
\label{eqb}
\ ds^{2}=\frac{L^2 S(z)}{z^2}[-g(z)dt^2+dx_{1}^{2}+e^{B^2 z^2}(dx_{2}^{2}+dx_{3}^{2})+\frac{dz^{2}}{g(z)}],
\end{equation}
where $S(z)$, $g(z)$ and $L$ denote scale factor, blackening function, and AdS length scale, respectively. In this model, the second gauge field is used to introduce a background magnetic field $B$ by $F_{(2)MN} = Bdx_2 \wedge dx_3$. The rotation symmetry $SO(3)$ is broken since the magnetic field $B$ is along the $x_1$-direction. It is worth mentioning that the background magnetic field $B$ is 5-dimensional (mass dimension one) in this metric. The physical and 4-dimensional magnetic field $\mathfrak{B}$ (mass dimension 2) could be obtained by rescaling $\mathfrak{B}\sim \frac{B}{L}$ \cite{DHoker:2009ixq,Dudal:2015wfn}. For our purpose of investigating the qualitative characteristics of a magnetic field intuitively, we use the 5-dimensional magnetic field $B$ in this work. In the Appendix, we solve the Einstein-Maxwell equations to justify the metric $Ans\ddot{a}tze$ for a small magnetic field, treating the $Ans\ddot{a}tze$ as a small perturbation to the known holographic solutions without a magnetic field.

The Einstein equations of motion can be obtained by using the $Ans\ddot{a}tze$ of Eq.(\ref{eqb})\cite{Bohra:2019ebj}
\begin{equation}
\label{eqc}
 \  g''(z)+g'(z) \left( 2 B^2 z + \frac{3S'(z)}{2S(z)} -\frac{3}{z}\right)-\frac{z^2 f_{1}(z)A'_{t}(z)^2 }{L^2 S(z)}= 0,
\end{equation}

\begin{equation}
\label{eqc}
 \  \frac{B^2 z e^{-2 B^2 z^2}f_{2}(z)}{L^2 S(z)}+2 B^2 g'(z)+g(z)\left(4 B^4 z +\frac{3B^2 S'(z)}{S(z)} - \frac{4 B^2}{z} \right)= 0,
\end{equation}

\begin{equation}
\label{eqd}
\  S''(z)- \frac{3S'(z)^2}{2S(z)}+\frac{2S'(z)}{z}+S(z)\left(\frac{4B^4 z^2}{3}+\frac{4B^2}{3} +\frac{1}{3} \phi'(z)^2 \right)= 0,
\end{equation}
and
\begin{equation}
\label{eqe}
\begin{split}
\ & \frac{g''(z)}{3g(z)} +\frac{S''(z)}{S(z)}+S'(z)\left(\frac{7B^2 z}{2S(z)}+\frac{3g'(z)}{2g(z)S(z)}-\frac{6}{z S(z)} \right)+ g'(z) \left(\frac{5B^2 z}{3g(z)}-\frac{3}{z g(z)} \right)\\
& +2B^4 z^2+ \frac{B^2 z^2 e^{-2 B^2 z^2}f_{2}(z)}{6L^2 g(z)S(z)}-6B^2 + \frac{2L^2 S(z)V(z)}{3z^2 g(z)}+\frac{S'(z)^2}{2S(z)^2} +\frac{8}{z^2} = 0.
\end{split}
\end{equation}

The equation of motion for the dilaton field is
\begin{equation}
\label{eqf}
\begin{split}
\ & \phi''(z)+\phi'(z)\left(2B^2 z +\frac{g'(z)}{g(z)}+\frac{3S'(z)}{2S(z)}-\frac{3}{z}\right)+\frac{z^2 A'_{t}(z)^2}{2L^2 g(z)S(z)}\frac{\partial f_{1}(\phi)}{\partial\phi} \\
& -\frac{B^2 z^2 e^{-2 B^2 z^2}}{2L^2 g(z)S(z)} \frac{\partial f_{2}(\phi)}{\partial\phi} -\frac{L^2 S(z)}{z^2 g(z)}\frac{\partial V(\phi)}{\partial\phi}= 0.
\end{split}
\end{equation}

The equation of motion for the first gauge field is given by
\begin{equation}
\label{eqg}
\  A''_t (z)+A'_t (z)\left(2 B^2 z +\frac{f'_{1}(z)}{f_{1}(z)}+\frac{S'(z)}{2S(z)}-\frac{1}{z}  \right)= 0.
\end{equation}

One can use the following boundary conditions to solve equations of motion
\begin{equation}
\label{eqh}
\begin{split}
& g(0)=1,\   g(z_h)=0,\\
& A_t(0)=\mu, \    A_t(z_h)=0,\\
& S(0)=1,\ \phi(0)=0,
 \end{split}
\end{equation}
where $z_h$ is the black hole horizon. The chemical potential $\mu$ can be obtained from the near boundary expansion of the zeroth component of the first gauge field.

The expression of the gauge field $A_t (z)$ is as follows:
\begin{equation}
\label{eqi}
A_t (z)=\mu [1-\frac{\int^{z}_0 d\xi \frac{\xi e^{-B^2 \xi^2}}{f_1(\xi)\sqrt{S(\xi)}}}{\int^{z_h}_0 d\xi \frac{\xi e^{-B^2 \xi^2}}{f_1(\xi)\sqrt{S(\xi)}}}]=\widetilde{\mu}\int^{z_h}_{z} d\xi \frac{\xi e^{-B^2 \xi^2}}{f_1(\xi)\sqrt{S(\xi)}}.
\end{equation}

The baryon density $\rho$ can be obtained from the expansion of the gauge field when it is close to the boundary in the holographic dictionary, $A_t=\mu- \rho z^2$. The baryon density in this model is
\begin{equation}
\label{eqi1}
\rho= \frac{\mu}{2 \int^{z_h}_0 d\xi \frac{\xi e^{-B^2 \xi^2}}{e^{-c \xi^2- B^2 \xi^2}}}.
\end{equation}

The forms of scale factor $S(z)$ and the gauge coupling function $f_1 (z)$ are
\begin{equation}
 \label{eqj}
\begin{split}
 & S(z) = e^{2P(z)},\\
 & f_1 (z) = \frac{e^{-cz^2 -B^2 z^2}}{\sqrt{S(z)}}.
 \end{split}
\end{equation}

Then the gauge coupling function $f_2 (z)$ is
\begin{equation}
 \label{eqj1}
 f_2 (z) = -\frac{L^2 e^{2 B^2 z^2 +2P(z)}}{z} \left[g(z) \left(4B^2 z+6P'(z)-\frac{4}{z}\right) +2g'(z) \right].
\end{equation}

The blackening function $g(z)$ is given as
\begin{equation}
 \label{eqk}
 g(z) = 1+\int^{z}_0 d\xi \xi^3 e^{-B^2 \xi^2-3P(\xi)}[K_1 + \frac{\widetilde{\mu}^2}{2c L^2}e^{c \xi^2}],
\end{equation}
with
\begin{equation}
 \label{eql}
 K_1 = -\frac{[1+ \frac{\widetilde{\mu}^2}{2c L^2}\int^{z_h}_0 d\xi \xi^3 e^{-B^2 \xi^2-3P(\xi)+c\xi^2}]}{\int^{z_h}_0 d\xi \xi^3 e^{-B^2 \xi^2-3P(\xi)}}.
\end{equation}

The dilaton field in terms of $P(z)$ is
\begin{equation}
 \label{eqm}
 \phi(z)= \int dz \sqrt{-\frac{2}{z} \left(3 z P''(z)-3zP'(z)^2 +6P'(z)+2 B^4 z^3+2B^2 z \right) }+K_2.
\end{equation}
where $K_2$ is used to ensure $\phi$ vanishes near the asymptotic boundary.

The potential of the dilaton field is given as
\begin{equation}
\label{eqn}
\begin{split}
\ & V(z)=\frac{g(z)}{L^2}\left(-\frac{9B^2 z^3 S'(z)}{2S(z)^2}+\frac{10B^2 z^2}{S(z)} -\frac{3z^2 S'(z)^2 }{S(z)^3}+\frac{12z S'(z)}{S(z)^2}+\frac{z^2 \phi'(z)^2}{2S(z)}-\frac{12}{S(z)}\right) \\
& -\frac{z^4 f_{1}(z)A'_{t}(z)^2}{2L^4 S(z)^2}+\frac{g'(z)}{L^2}\left(-\frac{B^2 z^3}{S(z)}-\frac{3z^2 S'(z)}{2S(z)^2}+\frac{3z}{S(z)} \right).
\end{split}
\end{equation}

The Hawking temperature and entropy in this background are
\begin{equation}
\label{eqo}
\begin{split}
& T=-\frac{z^3_h e^{-3P(z_h)-B^2 z^2_h}}{4\pi}[K_1 + \frac{\widetilde{\mu}^2}{2c L^2}e^{c z^2_h}],\\
& S= \frac{e^{B^2 z^2_h +3P(z_h)}}{4 z^3_h},
 \end{split}
\end{equation}
where Newton constant $G_5$ is set to be one.

From the above equations, one can find that these Einstein-Maxwell-dilaton gravity solutions are determined by $P(z)$ and the gauge kinetic function $f_{1}(z)$ (Eq.(\ref{eqj})). Different forms of $P(z)$ and $f_{1}(z)$ may lead to different gravity solutions.

We choose the following simple form of $P(z)$: \cite{Li:2017tdz}
\begin{equation}
\label{eqp}
 \ P(z)= -a\log(b z^2 +1).
\end{equation}

In this work, we want to study the phase structure of QCD matter for light quarks which is different from  that in \cite{Bohra:2019ebj}. One can set the parameter $c$ of the gauge kinetic function $f_{1}(z)$ to be $0.227$ when $B =0$ by matching with the mass spectrum of the $\rho$ meson with its excitations \cite{Yang:2014bqa}. In order to fit the confinement-deconfinement phase transition temperature at $B =0$ and $\mu =0$, we fix $a=3.943$ and  $b=0.0158$. The phase structure for light quarks with nonzero chemical potential with the form of Eq.(\ref{eqp}) has been studied in \cite{Li:2017tdz}. In this work, we extend the results to nonzero chemical potential and magnetic field cases. We discuss the effects of chemical potential and magnetic field on the location of the CEP simultaneously and study the effects of the magnetic field on the critical $\mu_{\scriptscriptstyle CEP}$ in the $T-\mu$ plane and the chemical potential on the critical $B_{\scriptscriptstyle CEP}$ in the $T-B$ plane.

\begin{figure}[H]
    \centering
      \setlength{\abovecaptionskip}{0.1cm}
    \includegraphics[width=14cm]{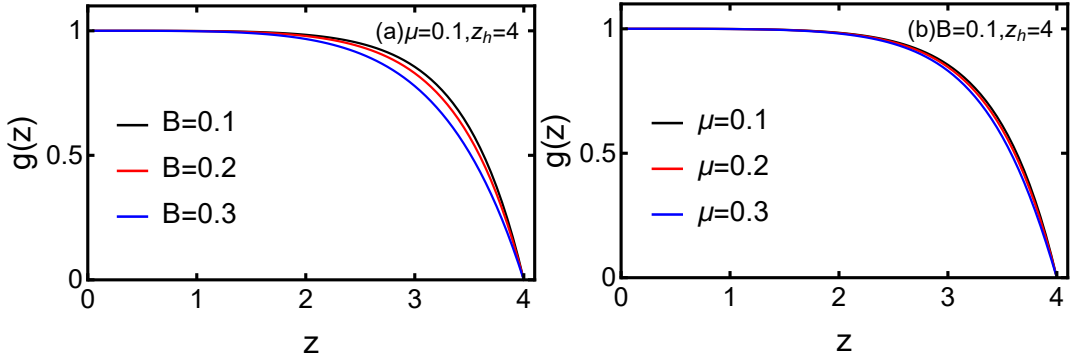}
    \caption{\label{fig1} $g(z)$ versus $z$. $B$ and $\mu$ are in units GeV.}
\end{figure}

\begin{figure}[H]
    \centering
      \setlength{\abovecaptionskip}{0.1cm}
    \includegraphics[width=14cm]{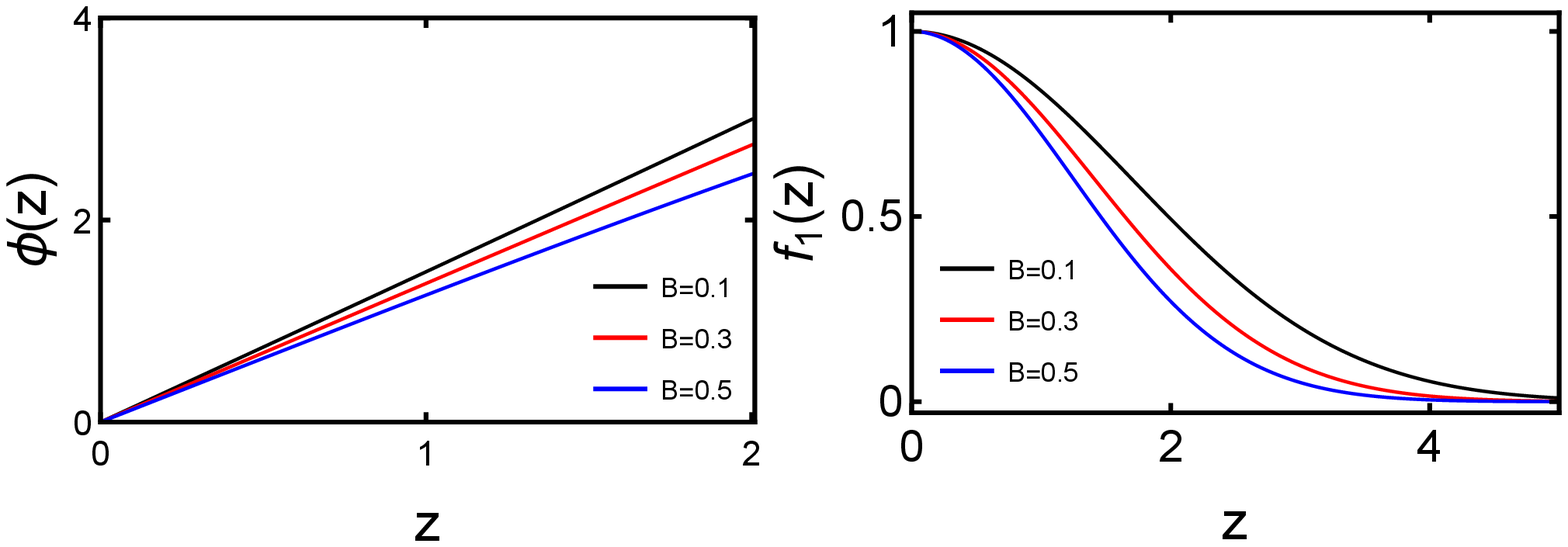}
    \caption{\label{fig2} $\phi(z)$ versus $z$ and gauge kinetic function $f_{1}(z)$ versus $z$. $B$ and $\mu$ are in units GeV.}
\end{figure}

\begin{figure}[H]
    \centering
      \setlength{\abovecaptionskip}{0.1cm}
    \includegraphics[width=14cm]{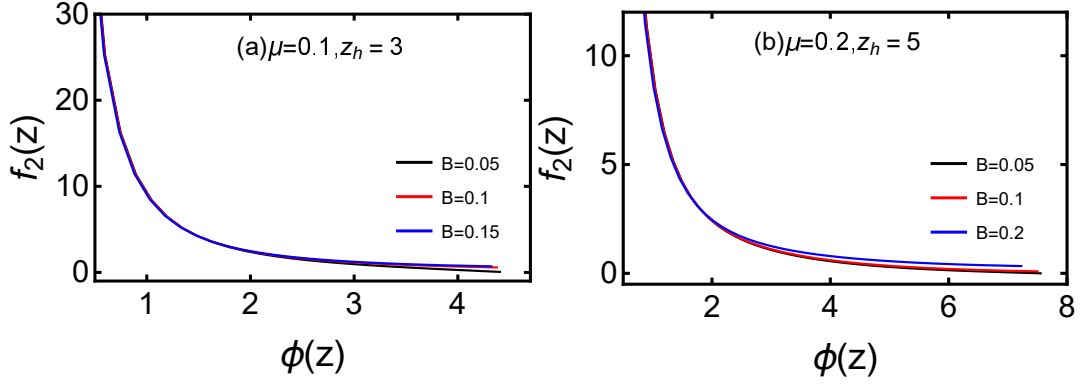}
    \caption{\label{fig3} Gauge kinetic function $f_{2}(z)$ versus $z$. $B$ and $\mu$ are in units GeV.}
\end{figure}

\begin{figure}[H]
    \centering
      \setlength{\abovecaptionskip}{0.1cm}
    \includegraphics[width=14cm]{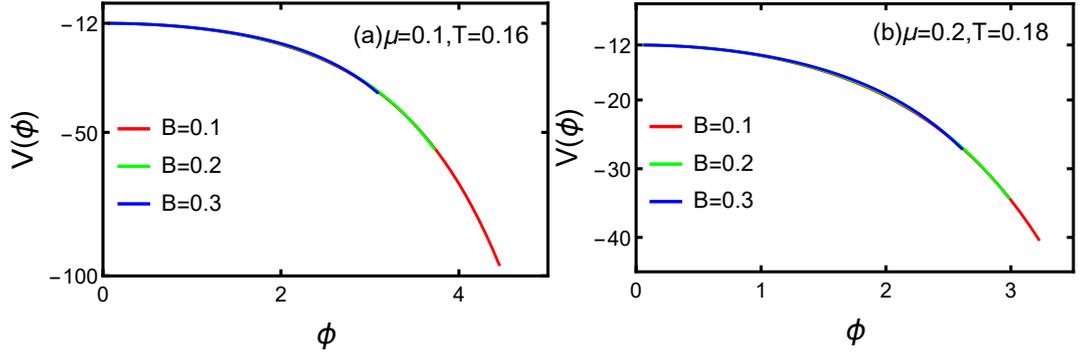}
    \caption{\label{fig4} $V(\phi)$ as a function of $\phi(z)$ with different values of magnetic field. $B$, $\mu$, and $T$ are in units GeV.}
\end{figure}

\begin{figure}[H]
    \centering
      \setlength{\abovecaptionskip}{0.1cm}
    \includegraphics[width=14cm]{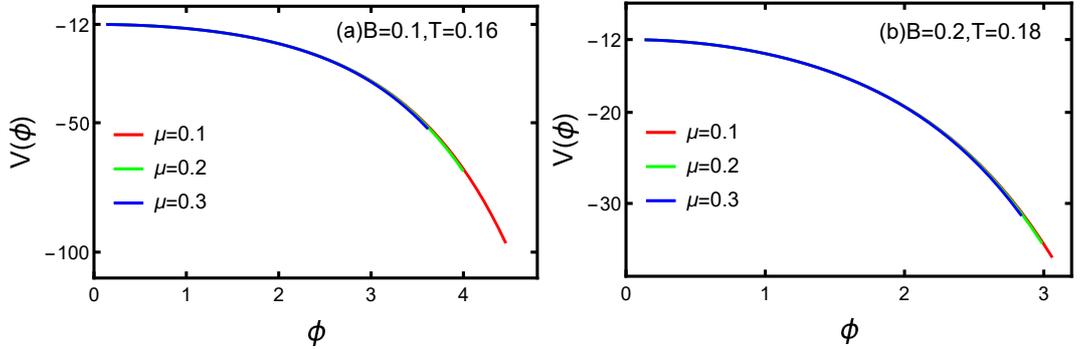}
    \caption{\label{fig5} $V(\phi)$ as a function of $\phi(z)$ with different values of chemical potential. $B$, $\mu$, and $T$ are in units GeV.}
\end{figure}

\begin{figure}[H]
    \centering
      \setlength{\abovecaptionskip}{0.1cm}
    \includegraphics[width=14cm]{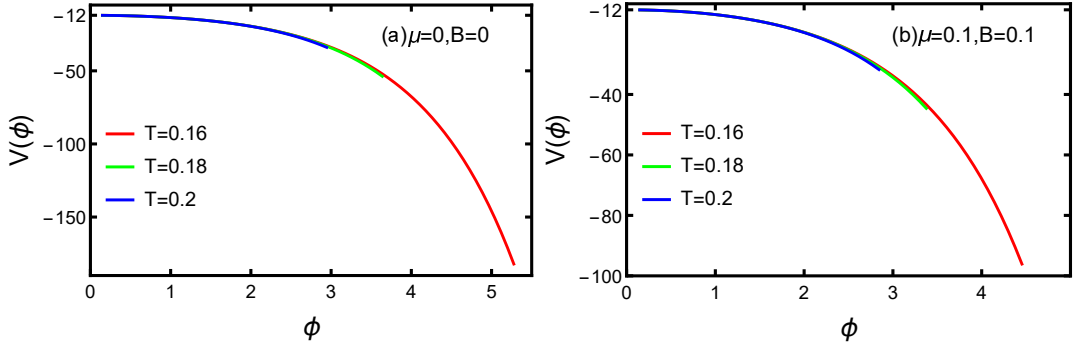}
    \caption{\label{fig6} $V(\phi)$ as a function of $\phi(z)$ with different values of temperatures. $B$, $\mu$, and $T$ are in units GeV.}
\end{figure}

Before studying the black hole thermodynamics and equations of state in this magnetized background, we want to check whether the metric solutions are self-consistent or not in Figs.~\ref{fig1} -~\ref{fig6}. If the metric solutions are self-consistent, the blackening function and dilaton field should satisfy the boundary condition of Eq.(9). The gauge kinetic functions $f_{1}$ and $f_{2}$ do not break the null energy condition (NEC); namely, they should be positive. The $B$, $\mu$, and $T$ dependence of $V(\phi)$ should be slight. We will check these in the following calculations. We take AdS length scale $L= 1$ in the calculations.

In Fig.~\ref{fig1}, we discuss the effects of the magnetic field and chemical potential on the blackening function $g(z)$. It is obvious that $g(0)=1$ and $g(z_h)=0$ with nonzero magnetic field and chemical potential which implies the blackening function satisfies the boundary condition. The blackening function decreases as the chemical potential and magnetic field increase.

In Fig.~\ref{fig2}, we plot $\phi(z)$ versus $z$ and $f_{1}(z)$ versus $z$ to study the effect of the magnetic field on the dilaton field and the gauge kinetic function. From Fig.~\ref{fig2} (a), we find $\phi(0)=0$ which implies the dilaton field satisfies the boundary condition. The dilaton field decreases as the magnetic field increases. The gauge kinetic function $f_{1}(z)$ is related to the mass spectrum of the $\rho$ meson. In \cite{Yang:2014bqa},  the authors fitted the mass spectrum of the $\rho(1^-)$ meson with $m^2_n=4 c n $, where $n$ is consecutive number and $c$ is the model parameter. It should be mentioned that $m^2_n$ represents the mass spectrum when the temperature, chemical potential and magnetic field vanish. Namely, the model parameter $c=0.227$ is fitted well with the mass spectrum at zero magnetic field. In Fig.2(b), we just want to check whether the gauge kinetic function $f_{1}$ is always positive or not at different $B$. The results show that $f_{1}(z)$ has no negative value and the null energy condition is not broken. Meanwhile, we observe that $f_{1}(z)$ decreases monotonically as $z$ increases.

In Fig.~\ref{fig3}, we plot the gauge kinetic function $f_{2}(z)$ versus $z$ with different magnetic fields. The gauge kinetic function $f_{2}(z)$ is calculated from equations of motion (EOMs). From the results, the magnetic field increases $f_{2}(z)$ slightly. The gauge kinetic function $f_{2}(z)$ decreases as dilaton field increases. Indeed, the values of $f_{2}(z)$ are close to zero near the horizon. From the results of the gauge kinetic functions $f_{1}(z)$ and $f_{2}(z)$, one can conclude that the values of the gauge kinetic function are non-negative which means $f_{1}(z)$ and $f_{2}(z)$ do not break the NEC.

In Figs.~\ref{fig4} -~\ref{fig6}, we study the effects of magnetic field, chemical potential, and temperature on the dilaton potential. After expanding the dialton field $\phi$ and dilaton potential $V(\phi)$ near the asymptotic boundary, one can rewrite $V(\phi)$ in terms of $\phi$, $V(\phi)= -12+\frac{m^2}{2}\phi^2+\cdot\cdot\cdot$. $m^2$ denotes the mass of $\phi$ and $m^2 =-3$ when $B=0$ which satisfies the Breitenlohner-Freedman (BF) bound \cite{Breitenlohner:1982jf}. Thus, $V(\phi)< -12$ when $ z\neq 0$ and $V(\phi)$ is bounded from above by its UV boundary value. This phenomenon can be observed from the figures. Also, $V(0)= -12 = 2\Lambda$ and $\Lambda$ denotes the cosmological constant in $AdS_{5}$.

We can find the dilaton potential is relevant with magnetic field, chemical potential, and temperature due to the metric and gauge field ansätze. We need to discuss the magnetic field, chemical potential, and temperature dependence of $V(\phi)$. If $V(\phi)$ is dependent on $B$, $\mu$, and $T$, then different values of $B$, $\mu$, and $T$ may lead to a different action and metric solution. In Figs.~\ref{fig4} -~\ref{fig6}, one can observe that the curves almost coincide together. Therefore, the $B$, $\mu$, and $T$ dependence of $V(\phi)$ is negligible. The back reaction of $B$ on action and metric solution is negligible.

\section{Thermodynamics in the holographic model}\label{sec:03}

In this section, we discuss the Hawking temperature and black hole free energy behaviors with the nonzero chemical potential and magnetic field in this holographic model. Then we plot the phase diagram in the $T-\mu$ and $T-B$ planes. The equations of state (pressure, entropy density, baryon density, specific heat, and sound speed) around phase transition temperature are studied in this section.
\subsection{Black hole thermodynamics}

\begin{figure}[H]
    \centering
      \setlength{\abovecaptionskip}{0.1cm}
    \includegraphics[width=14cm]{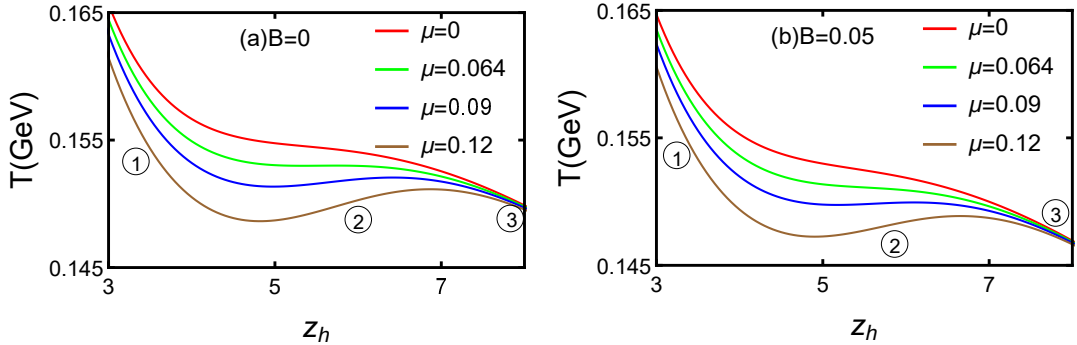}
    \caption{\label{fig7} Temperature $T$ as a function of horizon $z_{h}$ with different values of chemical potential. (a) for $B=0$ and (b) for $B=0.05$. $B$ and $\mu$ are in units GeV.}
\end{figure}

In Fig.~\ref{fig7}, we plot temperature $T$ as a function of horizon $z_{h}$ with different values of chemical potential. From the results of Fig.~\ref{fig7} (a), we find the temperature decreases monotonically as the horizon increases when $\mu= 0$. When the chemical potential is greater than the critical point ($\mu_c = 0.064GeV$ when $B= 0$), the local minimum value of temperature appears. In the $\mu\geq \mu_c$ case, there may exit three black hole phases. Specifically, the large black hole (\textcircled{1}), unstable phase (\textcircled{2}), and small black hole (\textcircled{3}) may exist simultaneously. The large/small black hole is thermodynamically stable while the unstable phase is not. This unstable feature leads to the negative values of specific heat and square of sound speed. From Fig.~\ref{fig7} (b), one can find the chemical potential dependence of temperature when $B=0.05$ is qualitatively the same as that when $B= 0$. Indeed, a phase transition may happen between the small black hole and large black hole.

The small-large black hole phase transition is dual to the confinement-deconfinement phase transition in \cite{He:2013qq}. However, the authors of \cite{Yang:2015aia} calculate the phase transition temperatures from black hole phases and open string configurations. They find the results have differences between the two methods. In further study of \cite{Dudal:2017max}, the authors consider the small-large black hole phase transition as the specious-confinement-deconfinement phase transition. Since the Polyakov loop expectation value is nonzero (extremely small), the small black hole phase is not exactly dual to confinement. The expectation value of the Polyakov loop is zero in the confinement phase. However, the Polyakov loop expectation value is nonzero (extremely small) in the small black hole phase from the results of \cite{Dudal:2017max}. The expectation value of the Polyakov loop only behaves as linear confinement for larger distances when $T$ is small. Therefore, the small black hole phase is called the specious-confinement phase.

From \cite{He:2020fdi}, the black hole free energy at fixed chemical potential, volume and magnetic field is
\begin{equation}
\label{eqn1}
 F=\int^\infty_{z_h} s \frac{dT}{d z_h}d z_h.
\end{equation}

\begin{figure}[H]
    \centering
      \setlength{\abovecaptionskip}{0.1cm}
    \includegraphics[width=15cm]{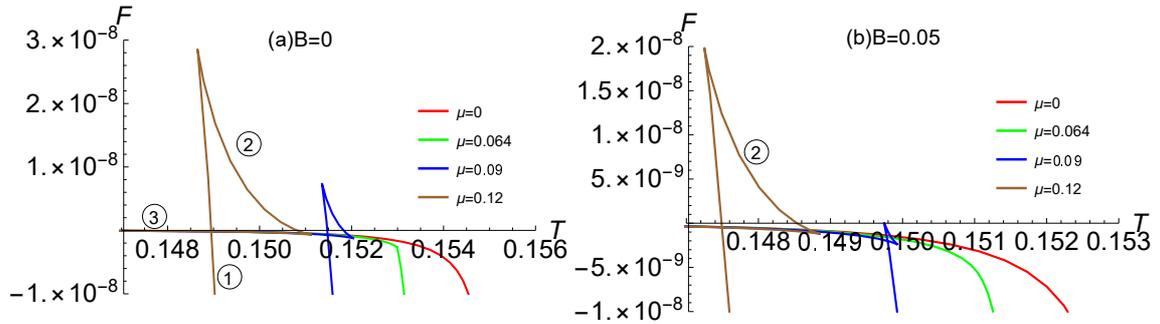}
    \caption{\label{fig8} Free energy $F$ as a function of temperature $T$ with different values of chemical potential. (a) for $B=0$ and (b) for $B=0.05$. $B$ and $\mu$ are in units GeV.}
\end{figure}
In Fig.~\ref{fig8}, we plot free energy $F$ as a function of temperature $T$ with different values of chemical potential. From the results of Fig.~\ref{fig8} (a), one can observe that the black hole free energy is a smooth function of temperature, which indicates the phase transition is a crossover with $\mu=0$. When $\mu\geq \mu_c$, the characteristic swallow-tailed shape emerges, which implies that the first-order phase transition happens. The small-large black hole phase transition happens at the kink of the swallow-tailed structure. From Fig.~\ref{fig8} (b), the results of the $B=0.05$ case are the same as those of the $B=0$ case. From Figs.~\ref{fig7} and ~\ref{fig8}, one finds that the chemical potential promotes crossover to first-order phase transition.

\begin{figure}[H]
    \centering
      \setlength{\abovecaptionskip}{0.1cm}
    \includegraphics[width=15cm]{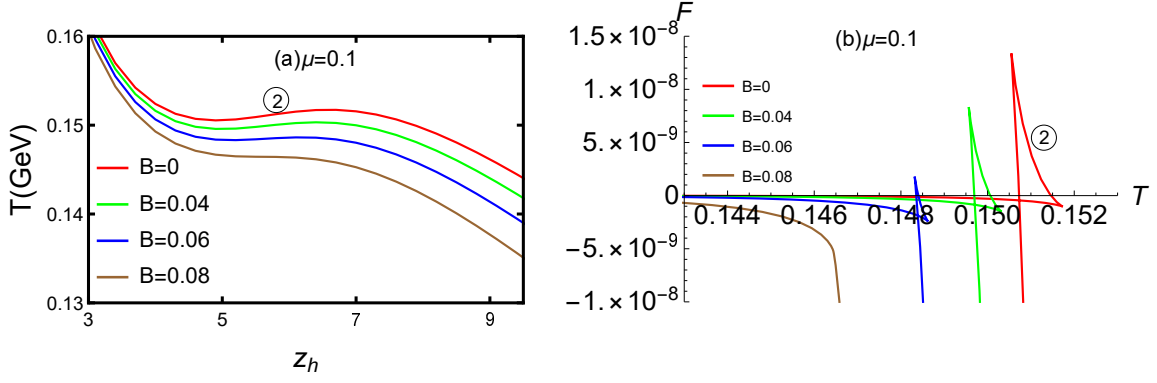}
    \caption{\label{fig9} Temperature $T$ versus horizon $z_{h}$ and free energy $F$ versus $T$ with different values of magnetic field when $\mu=0.1$. $B$ and $\mu$ are in units GeV.}
\end{figure}

In Fig.~\ref{fig9}, we study the effect of magnetic field on the temperature and black hole free energy. We plot temperature $T$ versus horizon $z_{h}$ and free energy $F$ versus $T$ with different values of magnetic field at finite chemical potential. Fig.~\ref{fig9} (a) indicates that the magnetic field suppresses the unstable phase (\textcircled{2}) which vanishes when the magnetic field is larger than a critical point $B_c$. The temperature decreases monotonically with horizon when $B\geq B_c$. From Fig.~\ref{fig9} (b), one finds that the characteristic swallow-tailed shape fades away with increasing magnetic field. When it reaches a critical point $B_c$, the free energy monotonously decreases with temperature. This means the magnetic field promotes the first-order phase transition to a crossover. Moreover, the phase transition temperature moves toward lower values with the increase of the magnetic field, implying an IMC.

\begin{figure}[H]
    \centering
      \setlength{\abovecaptionskip}{0.1cm}
    \includegraphics[width=14cm]{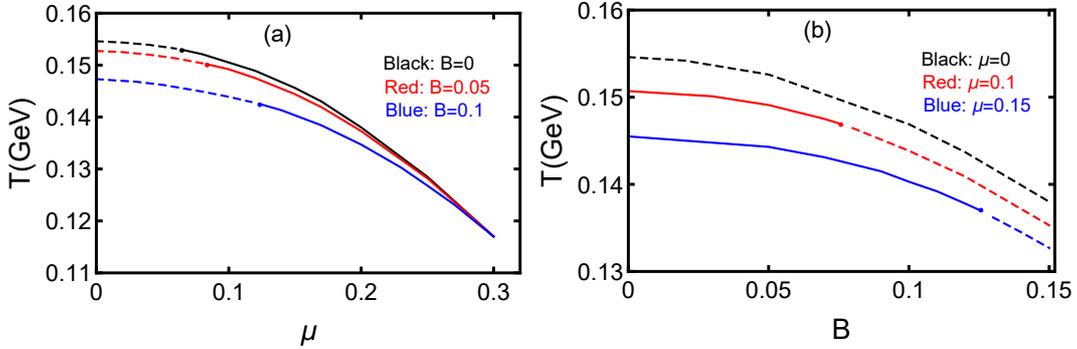}
    \caption{\label{fig10} The phase diagram in the $T-\mu$ plane and $T-B$ plane. The minima of sound speed squared $C^2_s$ are denoted by dashed lines. The first-order phase transitions (solid lines) are determined by black hole free energy. The black, red, and blue lines in the $T-\mu$ plane denote $B = 0, 0.05, 0.1GeV$, respectively. The locations of CEPs are $(\mu_{\scriptscriptstyle CEP},T_{\scriptscriptstyle CEP})=(0.064,0.153), (0.083,0.150),(0.123,0.1426)$ respectively. The black, red and blue lines in the $T-B$ plane denote $\mu = 0, 0.1, 0.15GeV$, respectively. The locations of CEP are $(B_{\scriptscriptstyle CEP},T_{\scriptscriptstyle CEP})=(0.075,0.147),(0.125,0.137)$ when $\mu = 0.1, 0.15GeV$, respectively. $B$ and $\mu$ are in units GeV.}
\end{figure}

In Fig.~\ref{fig10}, we plot the phase diagram in the $T-\mu$ and $T-B$ planes. Since the phase diagram for light quarks in $T-\mu$ plane is a smooth crossover at small chemical potential and turns into first-order at the CEP, we choose the minimum of the square of sound speed $C^2_s$ (dashed lines) to characterize the drastic change of degrees of freedom between QGP and the hadron phases in the crossover region. The first-order phase transition (solid lines) can be fixed by the free energy when the chemical potential is above a critical value $\mu_c$. Fig.~\ref{fig10} (a) shows the phase transition in the $T-\mu$ plane for $B = 0, 0.05, 0.1GeV$. The phase diagram shows crossover at $\mu < \mu_c$ and becomes first-order at $\mu$ $>$ $\mu_c$. The locations of the CEPs are $(\mu_{\scriptscriptstyle CEP},T_{\scriptscriptstyle CEP})=(0.064,0.153), (0.083,0.150),(0.123,0.1426)$ respectively. When increasing the magnetic field, the location of the CEP shifts toward the lower right region of the plane. It means the CEP would shift toward lower temperature and larger chemical potential when increasing $B$. The magnetic field increases critical $\mu_{\scriptscriptstyle CEP}$ in $T-\mu$ plane. We also find that the phase transition temperature when $\mu=0, B=0$ is in the $150-160 MeV$ region which is consistent with the lattice QCD prediction\cite{Borsanyi:2010bp,Borsanyi:2011sw}. Fig.~\ref{fig10} (b) shows the phase transition in the $T-B$ plane for $\mu = 0, 0.1, 0.15GeV$. One can find the obvious IMC. The locations of the CEPs are $(B_{\scriptscriptstyle CEP},T_{\scriptscriptstyle CEP})=(0.075,0.147),(0.125,0.137)$ when $\mu = 0.1, 0.15GeV$, respectively. In the $T-B$ plane and for zero $\mu$, the phase transition is always a crossover. In $T-B$ plane with finite $\mu$, the phase transition is of first-order at a small magnetic field and the CEP appears with the increasing magnetic field. The phase transition finally turns into a smooth crossover at a large magnetic field. When increasing the chemical potential, the location of the CEP shifts to the lower right plane. It means the CEP would shift toward lower temperature and larger magnetic field when increasing $\mu$. The chemical potential increases the critical $B_{\scriptscriptstyle CEP}$ of the CEP in the $T-B$ plane. One also can summarize that the chemical potential promotes the crossover to a first-order phase transition, while the magnetic field promotes the first-order phase transition to the crossover. It is worth mentioning that the phase diagram in the $T-\mu$ and $T-B$ planes for heavy quarks has been studied in \cite{He:2020fdi}.

\subsection{Equations of state}

\begin{figure}[H]
    \centering
      \setlength{\abovecaptionskip}{0.1cm}
    \includegraphics[width=15cm]{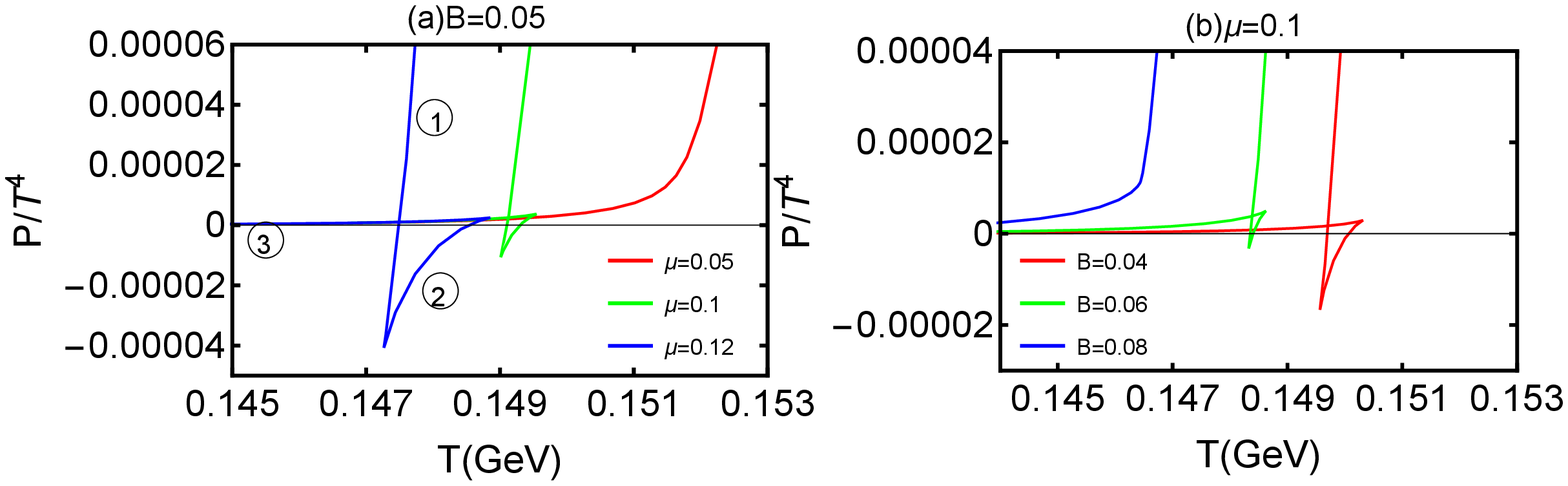}
    \caption{\label{fig11} The pressure $p/T^4$ versus temperature $T$ near the phase transition temperature. $B$ and $\mu$ are in units GeV.}
\end{figure}

In \cite{Ballon-Bayona:2022uyy}, the authors discuss the anisotropic pressures and sound speed. In this subsection, we focus on the pressure along the magnetic field and speeds of sound near the phase transition temperature with a parallel magnetic field.

The pressure $p$ along the $x_1$ direction (parallel to the magnetic field) is equal to $-F$. In Fig.~\ref{fig11}, we plot the pressure $p/T^4$ versus temperature $T$ near the phase transition temperature. From Fig.~\ref{fig11}(a), one can observe that the pressure is single-valued and always increases with the temperature when $0<\mu<\mu_c$. It indicates that the phase transition cross over in this region. When $\mu\geq \mu_c$, the unstable phase (\textcircled{2}) appears, which means the first-order phase transition happens. The findings of Fig.~\ref{fig11}(a) are consistent with the results of the $T-\mu$ phase diagram in Fig.~\ref{fig10}(a). These results also suggest that the chemical potential promotes the crossover into first-order phase transition which is consistent with the results of Fig.~\ref{fig7} and ~\ref{fig8}. From the results of Fig.~\ref{fig11}(b), the unstable phase appears at finite chemical potential and gradually disappears with increasing magnetic field which indicates the magnetic field causes the unstable phase to vanish. This finding is consistent with the results of Fig.~\ref{fig9}. It also shows that the phase transition finally turns into a smooth crossover from first-order at large magnetic field which agrees with the results of the $T-B$ phase diagram in Fig.~\ref{fig10}(b).

\begin{figure}[H]
    \centering
      \setlength{\abovecaptionskip}{0.1cm}
    \includegraphics[width=15cm]{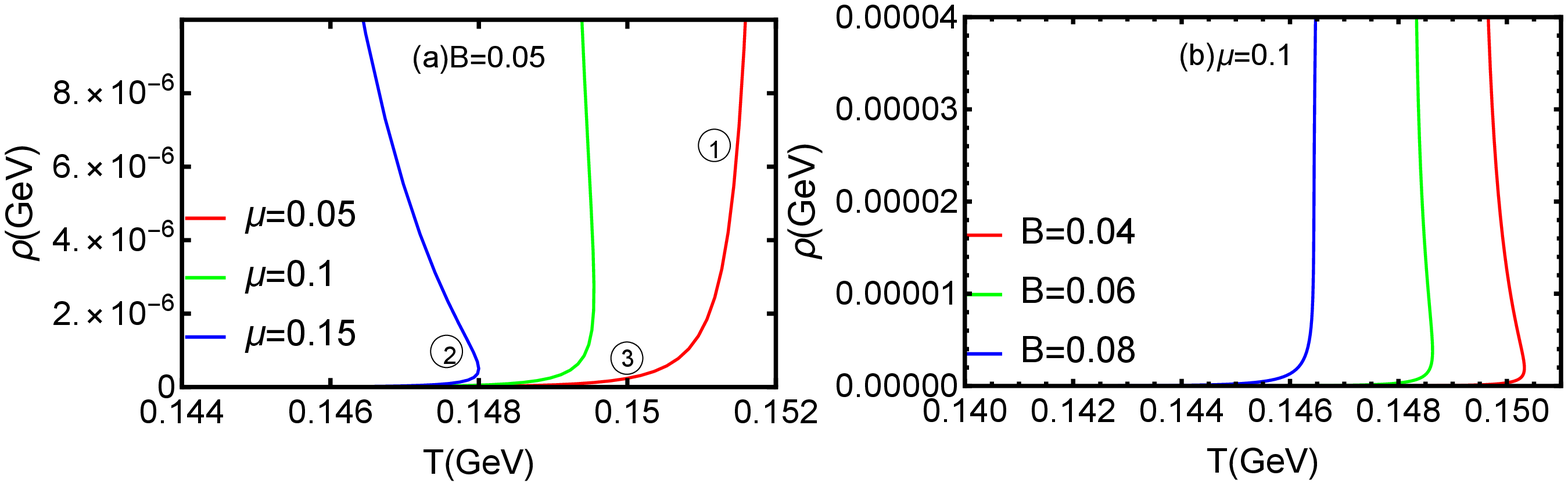}
    \caption{\label{fig12} The baryon density $\rho$ as a function of temperature $T$ near the phase transition temperature. $B$ and $\mu$ are in units GeV.}
\end{figure}

In Fig.~\ref{fig12}, we plot the baryon density $\rho$ as a function of temperature $T$ near the phase transition temperature. It is obvious that the baryon density is single-valued when $0<\mu<\mu_c$ while is multivalued around phase transition temperature when $\mu\geq \mu_c$ from Fig.~\ref{fig12}(a). It indicates that the phase transition is a crossover at $\mu < \mu_c$ and becomes first-order at $\mu$ $>$ $\mu_c$. The multivalued phenomenon becomes single-valued at large magnetic field in Fig.~\ref{fig12}(b), suggesting the phase transition finally turns into a smooth crossover from the first-order.

\begin{figure}[H]
    \centering
      \setlength{\abovecaptionskip}{0.1cm}
    \includegraphics[width=14cm]{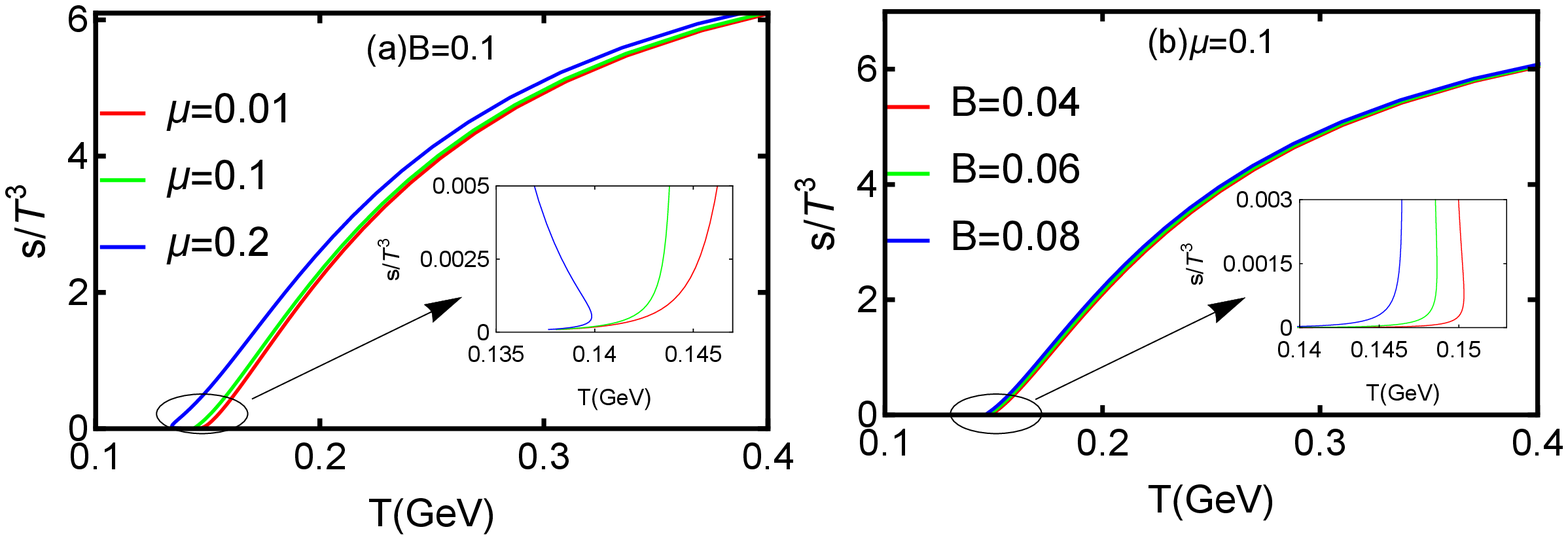}
    \caption{\label{fig13} The entropy density $s/T^3$ versus temperature $T$. $B$ and $\mu$ are in units GeV.}
\end{figure}

In Fig.~\ref{fig13}, we plot the entropy density $s/T^3$ versus temperature $T$. From Fig.~\ref{fig13} (a), one can find that the entropy is single-valued and always increases with the temperature when $0<\mu<\mu_c$ while is multivalued around the phase transition temperature when $\mu\geq \mu_c$. It indicates that the phase transition is a crossover at $\mu < \mu_c$ and becomes first-order at $\mu$ $>$ $\mu_c$. Interestingly, the magnetic field suppresses this multivalued phenomenon from Fig.~\ref{fig13}(b). The multivalued phenomenon turns into single-valued at a large magnetic field. The phase transition finally turns into a crossover from the first-order phase. Indeed, the magnetic field and chemical potential enhance the values of entropy.

\begin{figure}[H]
    \centering
      \setlength{\abovecaptionskip}{0.1cm}
    \includegraphics[width=14cm]{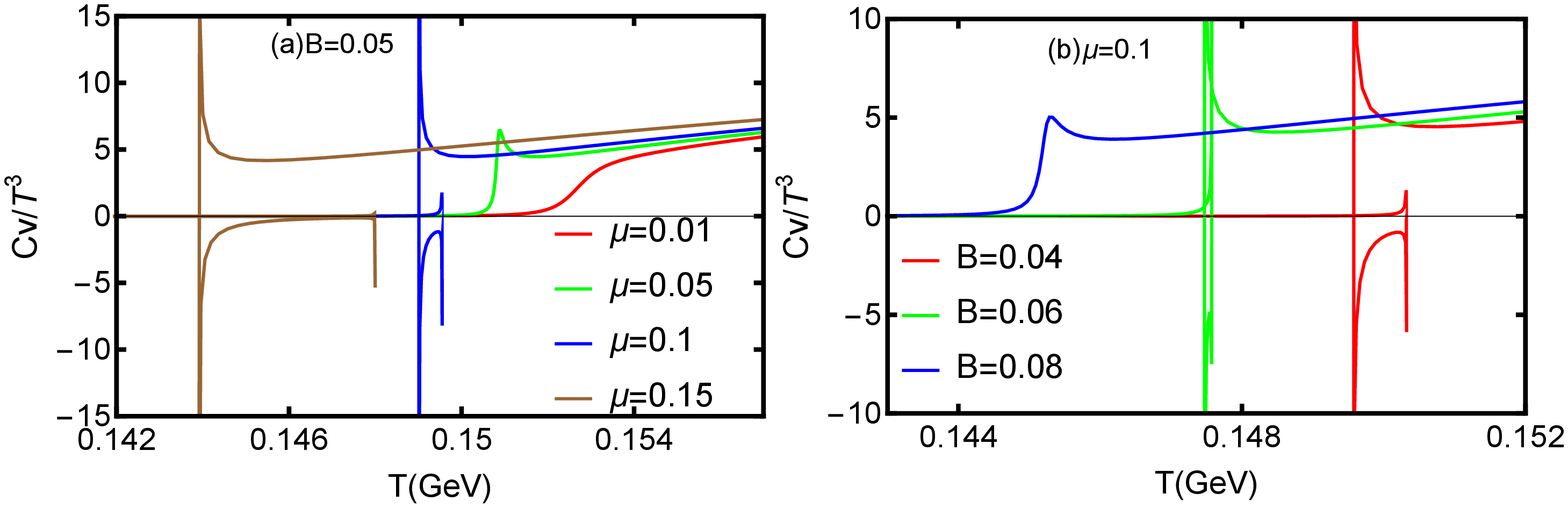}
    \caption{\label{fig14} The specific heat $C_V/T^3$ versus temperature $T$ near the phase transition temperature. $B$ and $\mu$ are in units GeV.}
\end{figure}

The specific heat is defined by
\begin{equation}
\label{eqo11}
 C_V= T (\frac{\partial s}{\partial T}).
\end{equation}

In Fig.~\ref{fig14}, we plot the specific heat $C_V/T^3$ versus temperature $T$. From Fig.~\ref{fig14} (a), one can find the specific heat is always positive when $0<\mu<\mu_c$, which means the black hole is thermodynamically stable. The unstable phase and the negative values of specific heat appear when $\mu\geq \mu_c$, signaling the emergence of a first-order phase transition. The results indicate the phase transition cross over at $\mu < \mu_c$ and becomes first-order at $\mu$ $>$ $\mu_c$. From Fig.~\ref{fig14}(b), we find that the unstable phenomenon gradually disappears with the increasing magnetic field and the specific heat is always positive at a large magnetic field. It indicates that the phase transition finally turns into a crossover.

\begin{figure}[H]
    \centering
      \setlength{\abovecaptionskip}{0.1cm}
    \includegraphics[width=14cm]{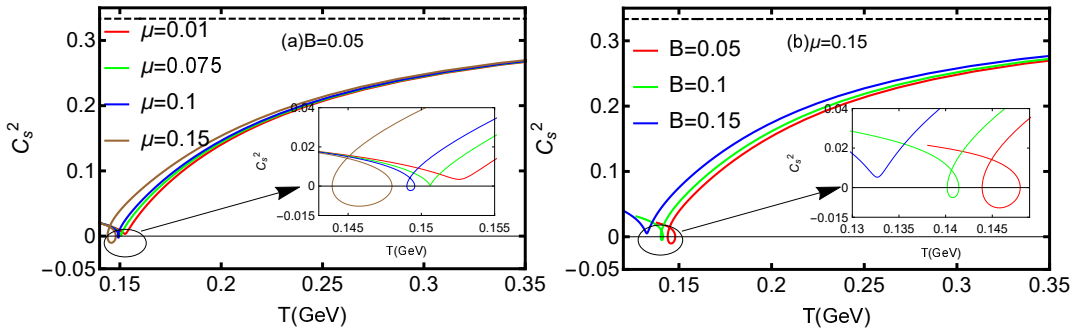}
    \caption{\label{fig15} The square of sound speed $C^2_s$ versus temperature $T$. The dotted black line denotes $C^2_s = 1/3$. $B$ and $\mu$ are in units GeV.}
\end{figure}

The square of sound speed along the $x_1$ direction is
\begin{equation}
\label{eqp11}
 C^2_s= \frac{\partial \ln T}{\partial \ln s}= \frac{s}{C_V}.
\end{equation}

In Fig.~\ref{fig15}, we plot square of sound speed  $C^2_s$ versus temperature $T$. From Fig.~\ref{fig15}(a), one observes that the square of sound speed is always positive when $0<\mu<\mu_c$, suggesting the black hole is thermodynamically stable. The negative values appear when $\mu\geq \mu_c$,  indicating thermodynamical instability. From Fig.~\ref{fig15}(b), we find that the unstable phenomenon fades away and the square of sound speed is always positive at large $B$. The results of $C^2_s$ are similar to the findings of $C_V/T^3$. Indeed, the chemical potential and magnetic field enhance the sound speed slightly at high temperatures.

We can find that the equations of state near the phase transition temperature are nonmonotonic and nontrivial. From the results of Figs.~\ref{fig11}-~\ref{fig15}, one finds the pressure, baryon density, and entropy density are single-valued and the specific heat and square of sound speed are always positive when $0<\mu<\mu_c$. When $\mu\geq \mu_c$, the pressure, baryon density, and entropy density are multivalued and the specific heat, and square of sound speed have negative values. It indicates the phase transition is a crossover at $\mu < \mu_c$ and becomes first-order at $\mu$ $>$ $\mu_c$ which is consistent with the results of the $T-\mu$ phase diagram in Fig.~\ref{fig10}(a). These results also suggest that the chemical potential promotes the crossover into first-order phase transition which is consistent with the results of Figs.~\ref{fig7} and ~\ref{fig8}.

At finite chemical potential, the multivalued phenomenon of the pressure, baryon density, and entropy density gradually disappears with increasing magnetic field and finally turns into single-valued at large magnetic field. The negative value phenomenon of the specific heat and square of sound speed fades away with increasing magnetic field. The specific heat and square of sound speed are always positive at a large magnetic field. It means the phase transition finally turns into a smooth crossover from first-order at a large magnetic field, which agrees with the results of the $T-B$ phase diagram in Fig.~\ref{fig10}(b). These results also indicate that the magnetic field promotes the first-order phase transition into crossover which is consistent with the results of Fig.~\ref{fig9}. The results of the equations of state can characterize the phase transition. We expect that the nontrivial behavior of the equations of state near the phase transition temperature could provide some theoretical reference for the study of QCD phase diagrams.

\section{Free energy of a $Q\bar{Q}$ pair in the holographic model}\label{sec:04}
In this section, we discuss the free energy of a quark-antiquark ($Q\bar{Q}$) pair in this holographic QCD model and compare it with the results of lattice QCD. The free energy of a $Q\bar{Q}$ pair is related to the on-shell action of a fundamental string from holography and can be deduced from the Nambu-Goto action. We study the anisotropic free energy since the rotation symmetry $SO(3)$ is broken by the magnetic field. We consider the $Q\bar{Q}$ pair is perpendicular and parallel to the magnetic field. From holography, the string world sheet action is calculated by the Nambu-Goto action for test string. Thus, one should transfer the metric to string frame from Einstein frame.

In the perpendicular case, the coordinates in Eq.(\ref{eqb}) are parameterized by
\begin{equation}
\label{eq21}
\ t=\tau,\quad x_{3}=\sigma,\quad  x_{1}=x_{2}=0,\quad z=z(\sigma).
\end{equation}

One can get the Lagrangian density from the Nambu-Goto action
\begin{equation}
\label{eq22}
\mathcal{L}=\frac{e^{2P_s(z)}}{z^2} \sqrt{e^{B^2 z^2}g(z)+\dot{z}^{2}},
\end{equation}
where $P_s (z)=P(z)+ \sqrt{\frac{1}{6}} \phi(z)$ in the string frame.

One obtains the interdistance $x_{\perp}$ of the $Q\bar{Q}$ pair perpendicular to the magnetic field
\begin{equation}
\label{eq24}
\ x_{\bot}=2 \int^{z_c}_{0} dz \sqrt{\frac{ z^4 e^{4P_s(z_c)} e^{B^2 z^2_c} g(z_c)}{z^4_c e^{4P_s(z)} e^{B^4 z^4}g^2(z)-z^4 e^{4P_s(z_c)} g(z)g(z_c)e^{B^2 z^2}e^{B^2 z^2_c}}},
\end{equation}
where $z_{c}$ is the tip of U-shaped string.

The free energy of the $Q\bar{Q}$ pair of the connected string in the perpendicular case is
\begin{equation}
\label{eq17}
\ F_{Q \bar{Q}(\perp)}= \frac{\sqrt{\lambda}}{\pi}\int^{z_c}_{0} dz \sqrt{\frac{ z^4_c e^{8P_s(z)} e^{B^2 z^2}g(z)}{z^4 z^4_c e^{4P_s(z)}e^{B^2 z^2} g(z)- z^8 e^{4P_s(z_c)} e^{B^2 z^2_c}g(z_c)}}.
\end{equation}
where $\sqrt{\lambda}=L^2/\alpha'$ represents the 't Hooft coupling.

The free energy of $Q\bar{Q}$ pair of the disconnected solution in the perpendicular case is
\begin{equation}
\label{eq171}
\ F_{discon(\perp)}= \frac{\sqrt{\lambda}}{\pi}\int^{z_h}_{0} dz \frac{e^{2P_s(z)}}{z^2}.
\end{equation}

In the parallel case, the interdistance $x_{\parallel}$ of the pair is
\begin{equation}
\label{eq28}
\ x_{\parallel}=2 \int^{z_c}_{0} dz \sqrt{\frac{ z^4 e^{4P_s(z_c)} g(z_c)}{z^4_c e^{4P_s(z)} g^2(z)-z^4 e^{4P_s(z_c)} g(z)g(z_c)}}.
\end{equation}

The free energy of the $Q\bar{Q}$ pair of the connected string in the parallel case is
\begin{equation}
\label{eq29}
\ F_{Q \bar{Q}(\parallel)}= \frac{\sqrt{\lambda}}{\pi}\int^{z_c}_{0} dz \sqrt{\frac{ z^4_c e^{8P_s(z)} g(z)}{z^4 z^4_c e^{4P_s(z)}g(z)- z^8 e^{4P_s(z_c)} g(z_c)}}.
\end{equation}

The free energy of the $Q\bar{Q}$ pair of the disconnected solution in the parallel case is
\begin{equation}
\label{eq172}
\ F_{discon(\parallel)}= \frac{\sqrt{\lambda}}{\pi}\int^{z_h}_{0} dz \frac{e^{2P_s(z)}}{z^2}.
\end{equation}

\begin{figure}[H]
    \centering
      \setlength{\abovecaptionskip}{0.1cm}
    \includegraphics[width=7.5cm]{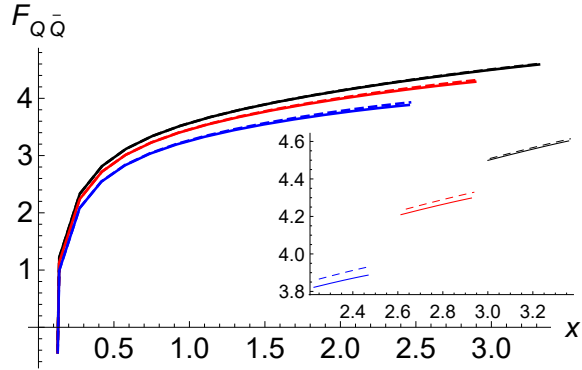}
    \caption{\label{fig16} The free energy of the $Q\bar{Q}$ pair as a function of interdistance $x$ in the large black hole when T=0.18 \ GeV and $\mu$=0.1 \ GeV. The solid line (dashed line) represents the $Q\bar{Q}$ pair is parallel (perpendicular) to the magnetic field. The black line, red line, and blue line represent B = 0.1, 0.2, 0.3 \ GeV respectively. We set $\lambda = 1$ in this figure.}
\end{figure}

In Fig.~\ref{fig16}, we plot the free energy of the $Q\bar{Q}$ pair as a function of interdistance $x$ in the large black hole when $T=0.18$ GeV and $\mu=0.1$ GeV. We observe that the free energy behaves as a Cornell-type potential. We can fit the free energy as $F=-\frac{0.2449}{x}+0.4703 x +3.266$ when $\mu=0.1$ GeV and $B = 0$. Furthermore, the magnetic field suppresses the free energy and this suppression is stronger when the $Q\bar{Q}$ pair is parallel to the magnetic field compared with the perpendicular case. This observation is consistent with the lattice QCD results \cite{Bonati:2016kxj}.

\begin{figure}[H]
    \centering
      \setlength{\abovecaptionskip}{0.1cm}
    \includegraphics[width=14cm]{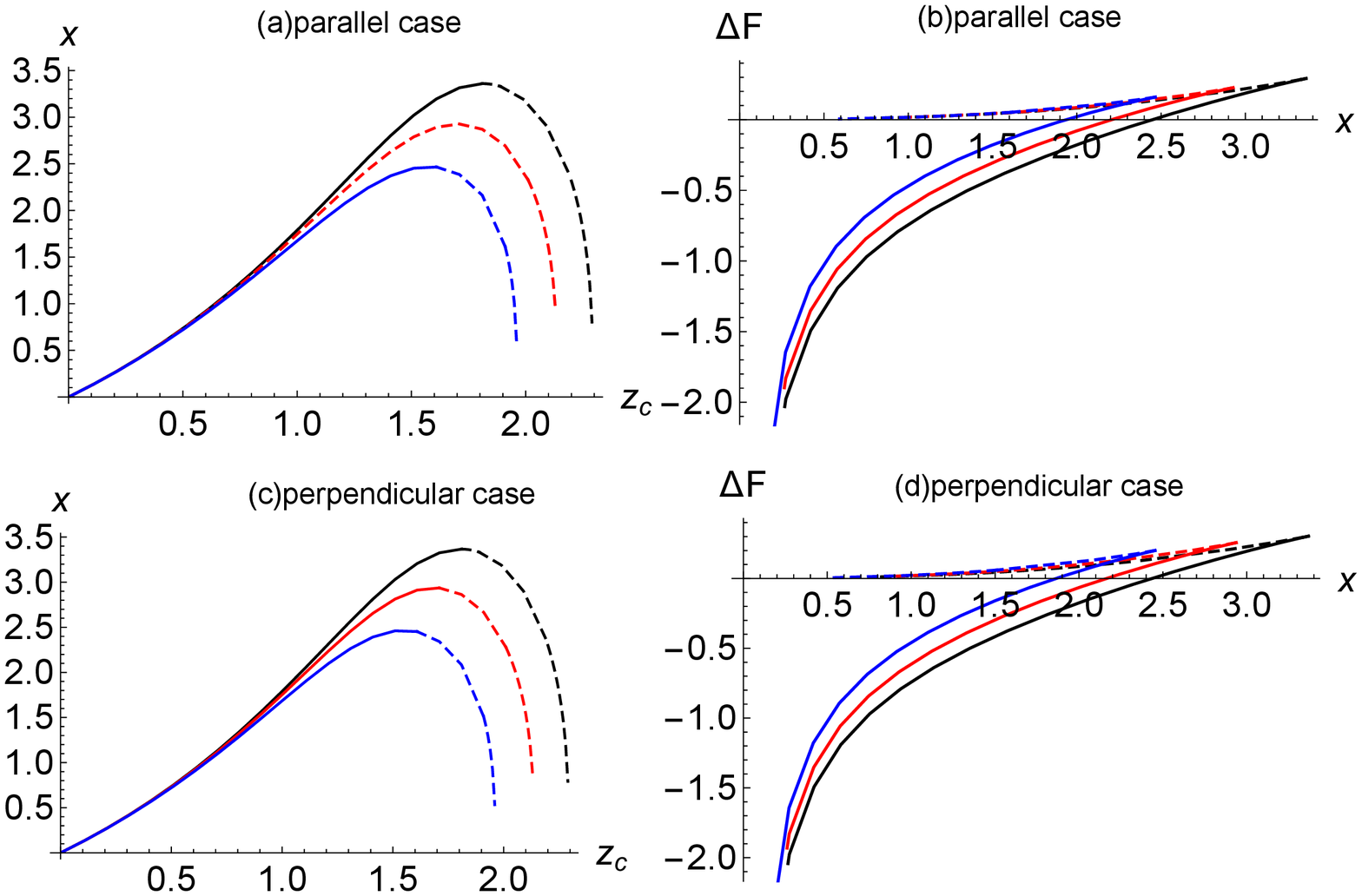}
    \caption{\label{fig17} The interdistance $x$ as a function of $z_c$ and $\Delta F= F_{Q \bar{Q}}- F_{discon}$ as a function of $x$ in the large black hole when T=0.18 \ GeV and $\mu$=0.1 \ GeV. The black line, red line and blue line represent B = 0.1, 0.2, 0.3 \ GeV respectively. Panels (a) and (b) represent the $q\bar{q}$ pair parallel to the magnetic field, while panels (c) and (d) represent the perpendicular case.  We set $\lambda = 1$ in this figure.}
\end{figure}

In Fig.~\ref{fig17}, we plot interdistance $x$ as a function of $z_c$ and $\Delta F= F_{Q \bar{Q}}- F_{discon}$ as a function of $x$ in the large black hole when T=0.18 \ GeV and $\mu$=0.1 \ GeV. In Figs.~\ref{fig17} (a) and (c), one can find that there exists a $x_{max}$ value above which the connected string disappears. Namely, the connected string configuration exists for small $z_c$ (solid lines) while the disconnected string arises for large $z_c$ (dotted lines). Moreover, one can find that the magnetic field decreases $x_{max}$ and promotes dissociation of the $Q\bar{Q}$ pair.

In Figs.~\ref{fig17} (b) and (d), we show the difference in free energy between the connected and disconnected string as $\Delta F= F_{Q \bar{Q}}- F_{discon}$. The solid and dotted lines in panels (b) and (d) denote the smaller and larger branches of interdistance $x$ respectively. It is obvious that the free energy of the smaller branch is always less than that of the larger branch. Moreover, there exists a critical value $x_{crit}$. $\Delta F$ is negative for $x< x_{crit}$, which implies the free energy of the connected configuration is less than that of the disconnected string. $\Delta F$ is positive for $x> x_{crit}$ which indicates the disconnected string has a lower free energy. $\Delta F =0 $ indicates a phase transition from a connected to a disconnected string at $x_{crit}$. This means the $Q\bar{Q}$ pair dissociates to a free quark and antiquark at large $x_{crit}$. We also find that $x_{crit}$ decreases with increasing magnetic field. Thus, the magnetic field enhances the dissociation of the $Q\bar{Q}$ pair.

\section{Energy loss in the holographic model}\label{sec:05}

In this section, we study the energy loss of fast moving probes near the phase transition temperature in a baryon-dense, magnetized, and strongly interacting medium. We discuss the effects of magnetic field and chemical potential on the drag force and $\hat{q}$.

\subsection{Nonmonotonic jet quenching parameter in the holographic model}
The jet quenching parameter plays a crucial role in energy loss of partons and can be deduced through lightlike adjoint Wilson loops. Since the magnetic field is along the $x_{1}$ direction and breaks the rotation symmetry, we study the anisotropic jet quenching parameter with nonzero magnetic field and chemical potential. One can summarize the main deduced formulas of the jet quenching parameter from \cite{Liu:2006ug}. In this paper, we focus on the jet quenching parameter around phase transition temperature. We observe a nonmonotonic temperature dependence behavior of the jet quenching parameter  $\hat{q}$ in this holographic AdS/QCD model.

First, we study the jet moving parallel to the magnetic field $\hat{q}_{\parallel}$. In this case, the results with momentum broadening along the $x_{2}$ or $x_{3}$ direction are the same. Here we consider the case that the jet is moving along the $x_{1}$ direction with momentum broadening along the $x_{2}$ direction, namely $\hat{q}_{(\parallel,\perp)}$.

We rewrite metric (\ref{eqb}) in the string frame
\begin{equation}
\label{eqbb4}
\ ds^{2}= g_{tt} dt^2 + g_{xx_{1}} dx_{1}^{2}+ g_{xx_{2}}(dx_{2}^{2}+dx_{3}^{2})+g_{zz} dz^{2},
\end{equation}
where $g_{tt}= -\frac{L^2 e^{2P_s(z)}}{z^2} g(z)$, $g_{xx_{1}}=\frac{L^2 e^{2P_s(z)}}{z^2}$, $g_{xx_{2}}=\frac{L^2 e^{2P_s(z)}}{z^2} e^{B^2 z^2}$  and $g_{zz}=\frac{L^2 e^{2P_s(z)}}{z^2} \frac{1}{g(z)}$.

With the light cone coordinates
\begin{equation}
\label{eqb5}
\ dt=\frac{dx^{+} + dx^{-}}{\sqrt{2}}, \ dx_{1}=\frac{dx^{+} - dx^{-}}{\sqrt{2}}.
\end{equation}
 the metric (\ref{eqbb4}) becomes
\begin{equation}
\label{eqb6}
\begin{split}
 ds^{2}=  \frac{1}{2}g_{tt} (dx^{+} + dx^{-})^2 + \frac{1}{2}g_{xx_{1}}(dx^{+} - dx^{-})^2 + g_{xx_{2}}(dx_{2}^{2}+dx_{3}^{2})+g_{zz} dz^{2}.
 \end{split}
\end{equation}

Then we choose the static gauge coordinate
\begin{equation}
\label{eqb8}
\ x^{-}=\tau,\ x_{2}=\sigma,\  x^{+}=x_{3}=const,\ z=z(\sigma).
\end{equation}
and the metric (\ref{eqb6}) becomes
\begin{equation}
\label{eqb9}
\ ds^{2}= \frac{1}{2}(g_{tt}+g_{xx_{1}})d\tau^2 +(g_{xx_{2}}+g_{zz}\dot{z}^2)d\sigma^2,
\end{equation}
where $\dot{z}=\frac{dz}{d\sigma}$.

Finally, the jet quenching parameter $\hat{q}_{(\parallel,\perp)}$ is
\begin{equation}
\label{eqb10}
\ \hat{q}_{(\parallel,\perp)}= \frac{\sqrt{\lambda}}{\pi} \left(\int^{z_h}_{0} dz \frac{z^2}{e^{2P_s(z)}e^{B^2 z^2} \sqrt{g(z)-g^2(z)}} \right)^{-1},
\end{equation}

Then we discuss the jet quenching parameter when the jet is moving perpendicular to the magnetic field $\hat{q}_{\perp}$. There are two cases for $\hat{q}_{\perp}$ since the momentum broadening may occur in different directions. First, we consider the jet is moving along the $x_{2}$ direction with momentum broadening along the $x_{1}$ direction, namely $\hat{q}_{(\perp,\parallel)}$, which can be expressed as
\begin{equation}
\label{eqb15}
\ \hat{q}_{(\perp,\parallel)}= \frac{\sqrt{\lambda}}{\pi} \left(\int^{z_h}_{0} dz \frac{z^2}{e^{2P_s(z)}\sqrt{g(z)e^{B^2 z^2}-g^2(z)}} \right)^{-1}.
\end{equation}

Then we consider the jet is moving along the $x_{2}$ direction with momentum broadening along the $x_{3}$ direction, namely, $\hat{q}_{(\perp,\perp)}$, which is obtained as
\begin{equation}
\label{eqb16}
\ \hat{q}_{(\perp,\perp)}= \frac{\sqrt{\lambda}}{\pi} \left(\int^{z_h}_{0} dz \frac{z^2}{e^{2P_s(z)}e^{B^2 z^2} \sqrt{g(z)e^{B^2 z^2}-g^2(z)}} \right)^{-1}.
\end{equation}

\begin{figure}[H]
    \centering
      \setlength{\abovecaptionskip}{0.1cm}
    \includegraphics[width=14cm]{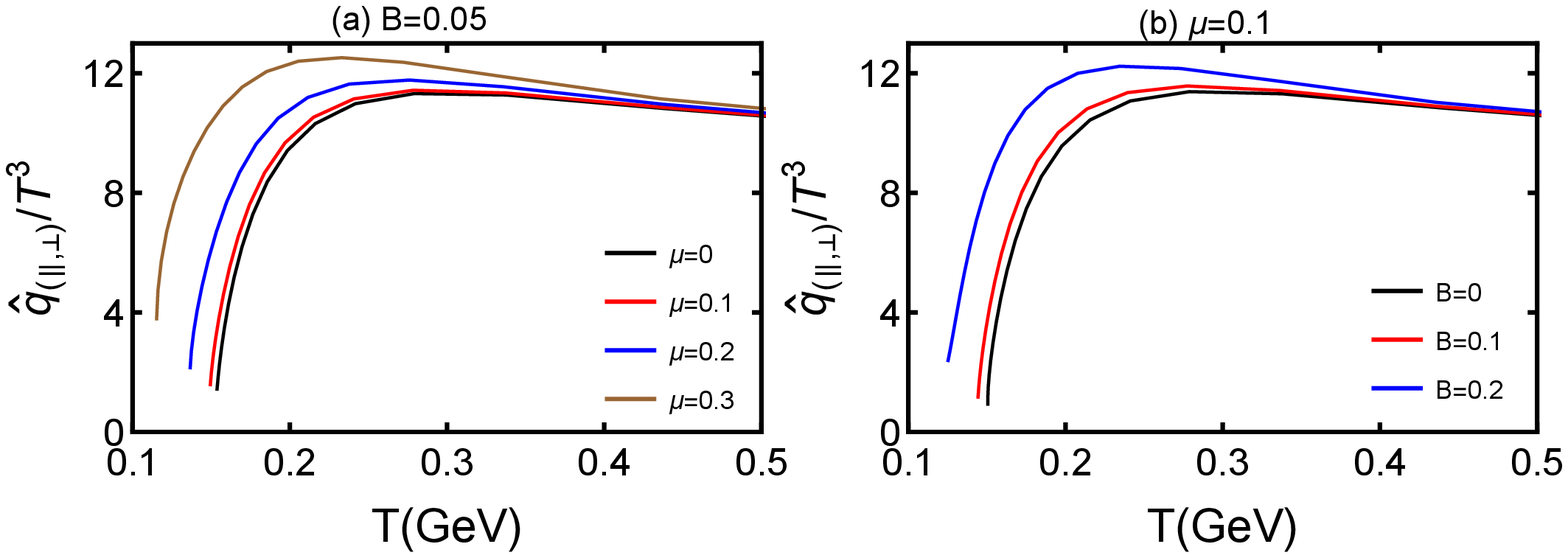}
    \caption{\label{fig18} $\hat{q}_{(\parallel,\perp)}/ T^3$ versus temperature $T$ when $\lambda$ = 1. $B$ and $\mu$ are in units GeV.}
\end{figure}

In Fig.~\ref{fig18}, we plot $\hat{q}_{(\parallel,\perp)}/ T^3$ versus temperature $T$. One can find that the magnetic field and chemical potential enhance the jet quenching parameter. With the critical temperature for deconfinement phase transition $T_c = 0.155GeV$ with $B=0, \mu=0$ in this model, $\hat{q}_{(\parallel,\perp)}/ T^3$ is temperature dependent and reaches a peak around $1.3T_c - 1.4T_c $ from the numerical results. Moreover, the peak value of $\hat{q}_{(\parallel,\perp)}/ T^3$ is moving toward lower temperature with increasing magnetic field or chemical potential. This phenomenon is consistent with the deconfinement phase transition temperature decrease with increasing $B$ or $\mu$.

\begin{figure}[H]
    \centering
      \setlength{\abovecaptionskip}{0.1cm}
    \includegraphics[width=14cm]{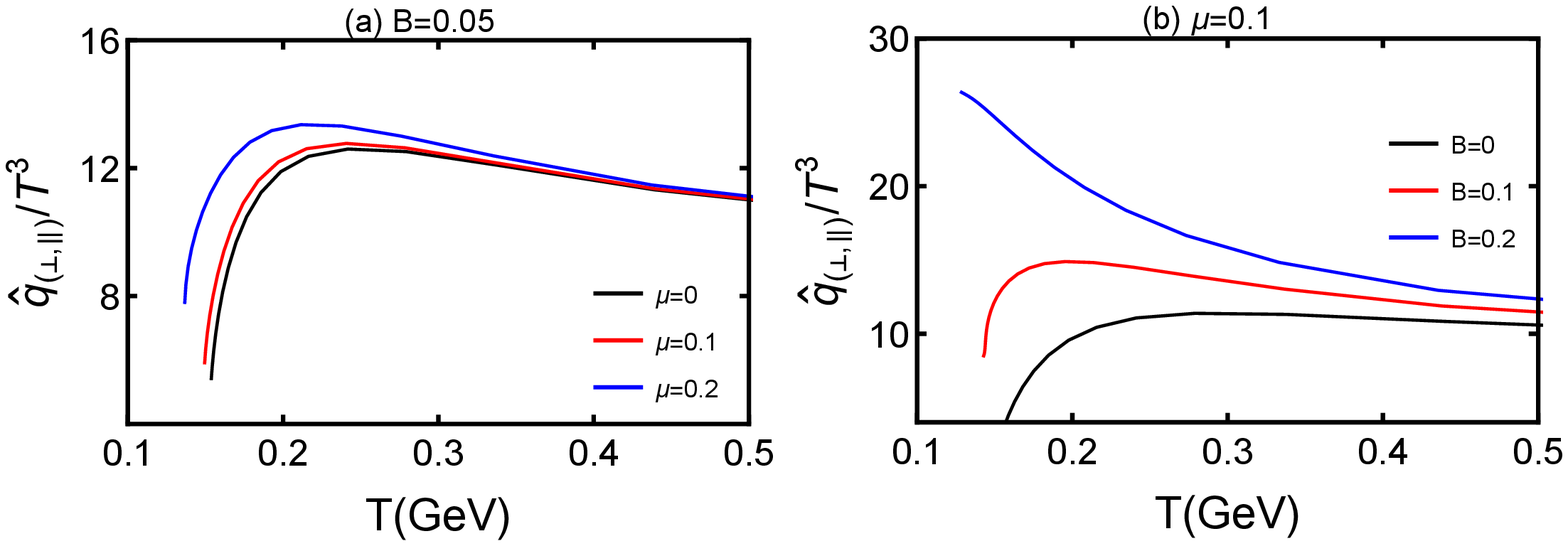}
    \caption{\label{fig19} $\hat{q}_{(\perp,\parallel)}/ T^3$ versus temperature $T$ when $\lambda$ = 1. $B$ and $\mu$ are in units GeV.}
\end{figure}
\begin{figure}[H]
    \centering
      \setlength{\abovecaptionskip}{0.1cm}
    \includegraphics[width=14cm]{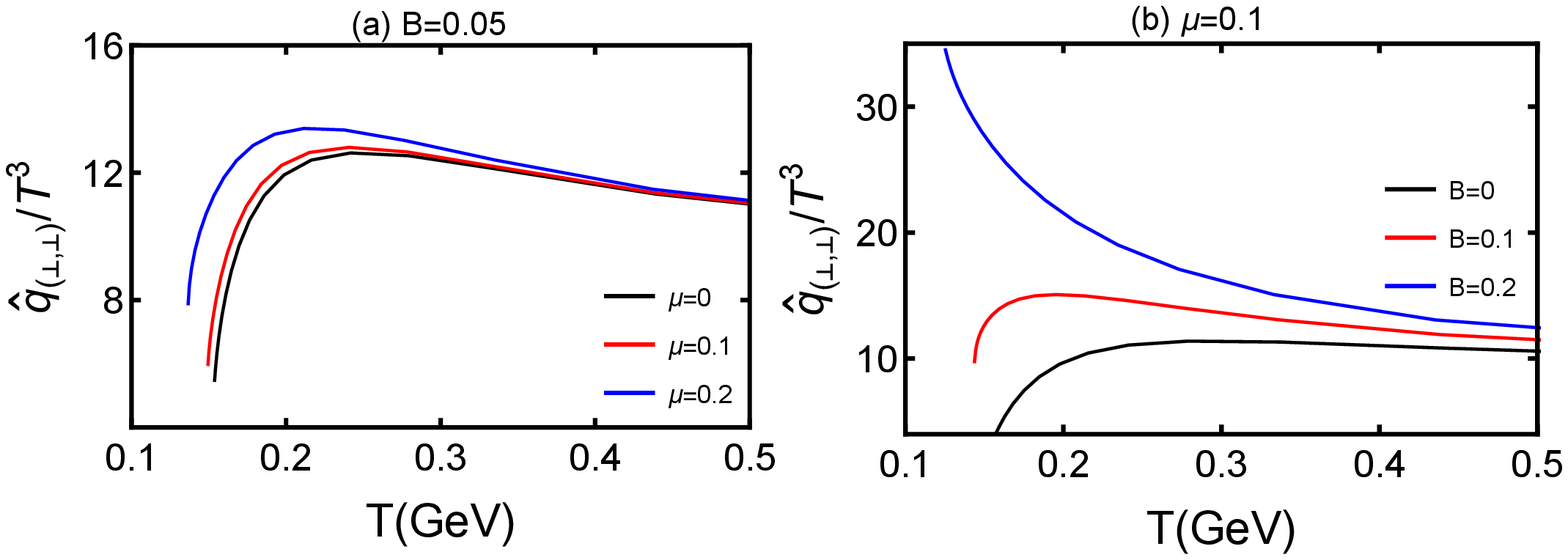}
    \caption{\label{fig20} $\hat{q}_{(\perp,\perp)}/ T^3$ versus temperature $T$ when $\lambda$ = 1. $B$ and $\mu$ are in units GeV.}
\end{figure}

We plot $\hat{q}_{(\perp,\parallel)}/ T^3$ and $\hat{q}_{(\perp,\perp)}/ T^3$ versus temperature $T$, respectively, in Figs.~\ref{fig19} and ~\ref{fig20}. We observe that the magnetic field and chemical potential enhance $\hat{q}_{(\perp,\parallel)}/ T^3$ and $\hat{q}_{(\perp,\perp)}/ T^3$. In cases of the jet moving perpendicular to the magnetic field cases, there is also a peak around $1.3T_c - 1.4T_c$ which implies $\hat{q}_{\perp}/ T^3$ has an enhancement around the phase transition temperature. One can summarize that $\hat{q}/ T^3$ is enhanced near the phase transition temperature which agrees with the lattice QCD results\cite{Kumar:2020wvb}. Moreover,  $\hat{q}_{\perp}/ T^3$ decreases monotonously with temperature at a large magnetic field.

In this section, we want to study the behaviors of $\hat{q}/ T^3$ near the first-order phase transition temperature. From the results of Figs.~\ref{fig7}-~\ref{fig9}, one can find that the metric solutions contain small black hole, large black hole, and unstable black hole phases simultaneously when the chemical potential is large or the magnetic field is small. In this case, the first-order phase transition exists. We study the behaviors of $\hat{q}/ T^3$ in the three phases simultaneously. We do not fix the temperature to be constant. Then one can find that $\hat{q}/ T^3$ is enhanced near the first-order phase transition. In Fig.~\ref{fig9}, the first-order phase transition disappears and crossover appears at large $B$. Thus, in Figs.~\ref{fig19}(b) and ~\ref{fig20}(b), $\hat{q}_{\perp}/ T^3$ decreases monotonously with temperature and there is no enhanced phenomenon at large $B$. Therefore, we can use the jet quenching parameter to characterize phase transition.

Moreover, the peak value is moving toward lower temperature with increasing magnetic field or chemical potential. This phenomenon is consistent with the deconfinement phase transition temperature decrease with increasing $B$ or $\mu$. From the discussions above, one can find that $\hat{q}/ T^3$ is enhanced near  the first-order phase transition. From the results of the phase diagram in Fig.10, one can find that the deconfinement phase transition temperature decreases with increasing $B$ or $\mu$. Thus, the peak value of $\hat{q}/ T^3$ is moving toward lower temperature with increasing $B$ or $\mu$.

\begin{figure}[H]
    \centering
      \setlength{\abovecaptionskip}{0.1cm}
    \includegraphics[width=14cm]{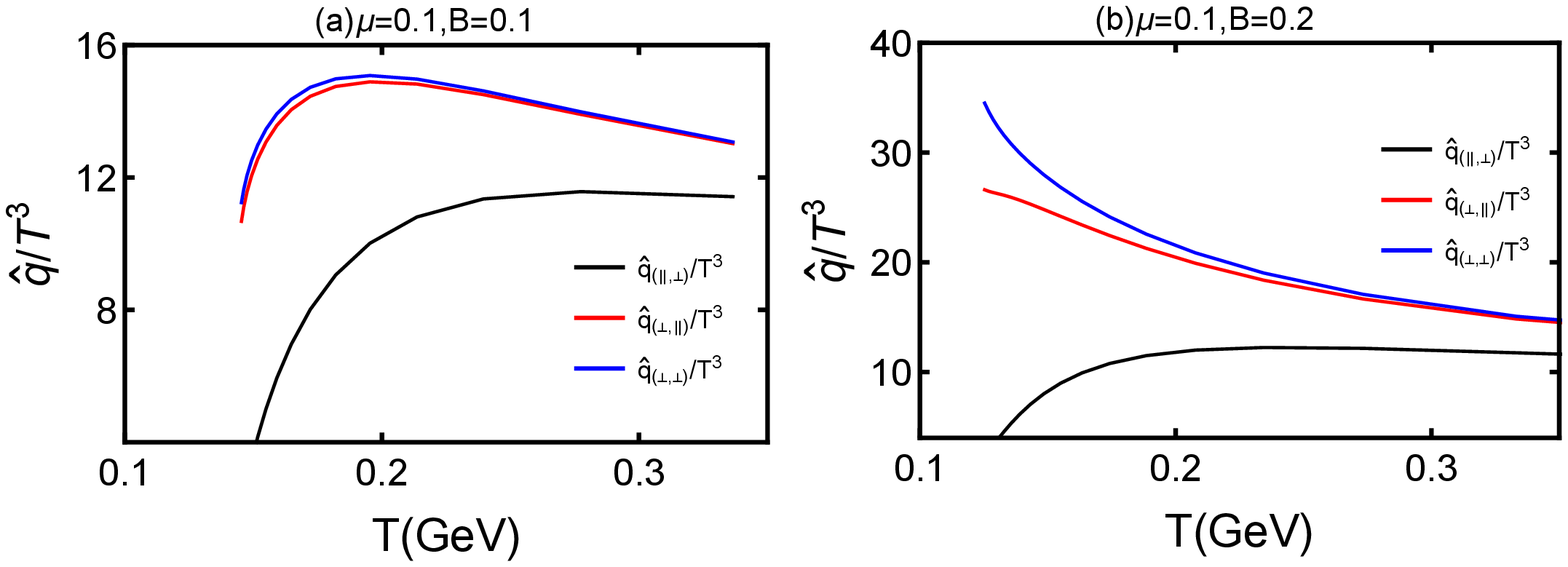}
    \caption{\label{fig21} $\hat{q}/ T^3$ versus temperature $T$ when $\lambda$ = 1. $B$ and $\mu$ are in units GeV.}
\end{figure}

\begin{table}[H]
 \begin{minipage} [t]{1.0\textwidth}
  \centering
       \begin{tabular}{c|c|c|c|c|c}
\hline
\diagbox{$\hat{q} $}{($\mu$,B)} & (0, 0) & (0, 0.1) & (0.1, 0.1) & (0.1, 0.2) & (0.2, 0.2) \\
\hline
$\hat{q}_{(\parallel,\perp)}$ & 6.71 & 6.82 &6.86 & 7.13 & 7.31\\
$\hat{q}_{(\perp,\parallel)}$ & 6.71& 8.00 & 8.05  &9.36  & 9.55\\
$\hat{q}_{(\perp,\perp)}$ & 6.71& 8.04 & 8.09  &9.55 & 9.75 \\
\hline
\end{tabular}
  \end{minipage}
\caption{\label{Table1} The values of $\hat{q} (GeV^2/fm)$ with different magnetic fields and chemical potentials when $T = 0.3$ GeV and $\lambda$ = 6$\pi$.}
\end{table}

We plot $\hat{q}/ T^3$ versus temperature $T$ in Fig.~\ref{fig21}. One can find $\hat{q}_{(\perp,\perp)} > \hat{q}_{(\perp,\parallel)} > \hat{q}_{(\parallel,\perp)}$, namely, $\hat{q}_{\perp}> \hat{q}_{\parallel}$. It indicates that the jet may lose more energy when it is moving perpendicular to the magnetic field. In order to compare with the RHIC data, we calculate the numerical results of the jet quenching parameter with $\lambda$ = 6$\pi$ and $T = 0.3$ GeV in Table.~\ref{Table1}. One finds that the jet quenching parameter $\hat{q}$ is in $6 - 10 GeV^2/fm$ region in this holographic model which agrees with the extracted values from RHIC data \cite{Edelstein:2008cp}.

\subsection{Nonmonotonic drag force in the holographic model}

From the description of the trailing string model \cite{Gubser:2006bz,Herzog:2006gh}, the heavy probe passing through the strongly coupled plasma in the spatial direction with a fixed velocity $\upsilon $ can be treated as the end point of an open string moving on the boundary. The rest of the string trails behind it and the loss of averaged momentum per unit time ($dp/dt$) can be holographically calculated by the energy flow ($dE/dx$) from the string end point into world sheet horizon.

The anisotropic drag force in the EMD model has been discussed in \cite{Finazzo:2016mhm}. We summarize the main deduced formulas of drag force. When the heavy quark moves parallel to the magnetic field in the $x_1$ direction, the drag force is

\begin{equation}
\label{eqc1}
\ f_\parallel = -\frac{1}{2\pi\alpha'} g_{xx_{1}}(z^{\parallel}_{c}) \upsilon,
\end{equation}
where $z^{\parallel}_{c}$ is the string world sheet horizon. $z^{\parallel}_{c}$ can be obtained by the following equation

\begin{equation}
\label{eqc2}
\ g_{tt}(z^{\parallel}_{c})=-g_{xx_{1}}(z^{\parallel}_{c})\upsilon^2.
\end{equation}

When the heavy quark moves perpendicularly to the magnetic field in the $x_2$ direction, the drag force is
\begin{equation}
\label{eqc3}
\ f_\perp = -\frac{1}{2\pi\alpha'} g_{xx_{2}}(z^{\perp}_{c}) \upsilon,
\end{equation}

and $z^{\perp}_{c}$ is determined by
\begin{equation}
\label{eqc4}
\ g_{tt}(z^{\perp}_{c})=-g_{xx_{2}}(z^{\perp}_{c})\upsilon^2.
\end{equation}

The drag force in $\mathcal{N} = 4$ SYM theory \cite{Gubser:2006bz} is
\begin{equation}
\label{eqz}
f_{\scriptscriptstyle SYM}= -\frac{\pi T^{2} \sqrt{\lambda}}{2} \frac{\upsilon}{\sqrt{1-\upsilon^{2}}},
\end{equation}
where $ \sqrt{\lambda}=\frac{L^{2}}{\alpha'}=\sqrt{g^{2}_{\scriptscriptstyle YM}N_{c}}$.

\begin{figure}[H]
    \centering
      \setlength{\abovecaptionskip}{0.1cm}
    \includegraphics[width=14cm]{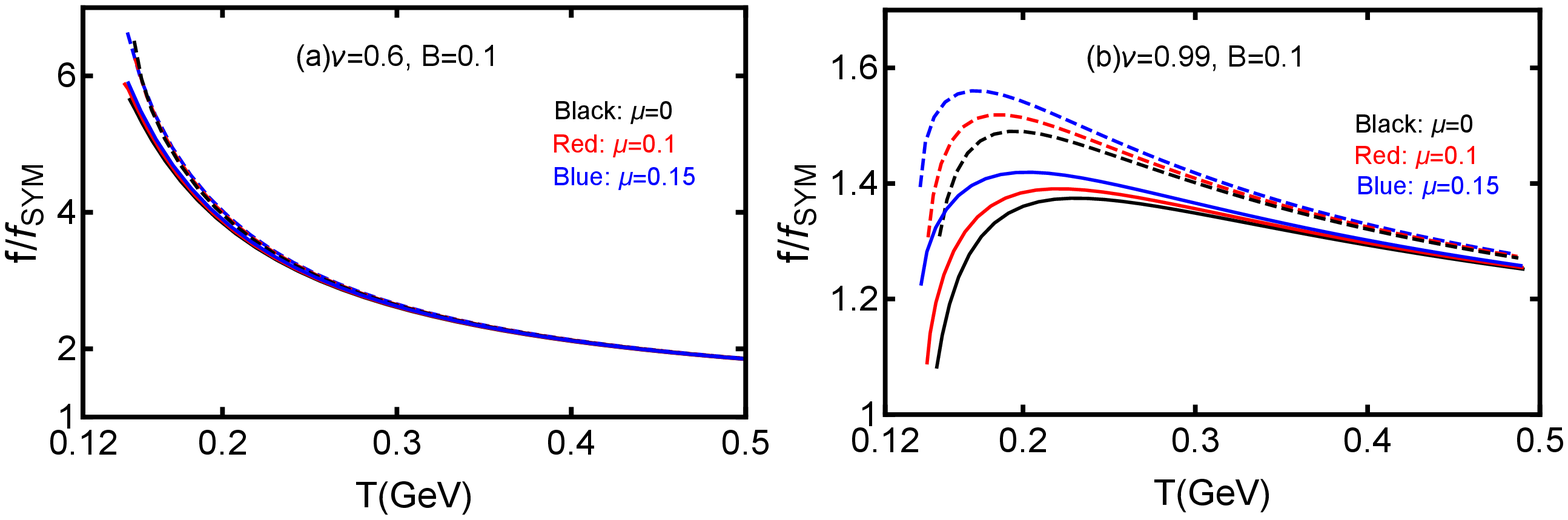}
    \caption{\label{fig22} Drag force normalized by its conformal limit $f/f_{\scriptscriptstyle SYM}$ versus temperature $T$ with different chemical potentials. The solid (dashed) line represents the heavy quark moving parallel (perpendicular) to the magnetic field. The black, red and blue lines denote $\mu = 0, 0.1, 0.15GeV$, respectively. $B$ and $\mu$ are in units GeV.}
\end{figure}

In Fig.~\ref{fig22}, we plot drag force normalized by its conformal limit $f/f_{\scriptscriptstyle SYM}$ versus temperature $T$ with different chemical potentials. Fig.~\ref{fig22}(a) is for moderate velocity $\upsilon = 0.6$, and Fig.~\ref{fig22}(b) is for ultrarelativistic velocity $\upsilon = 0.99$. One can observe that drag force with small velocity is not sensitive to the chemical potential of the medium since the curves are indistinguishable in Fig.~\ref{fig22}(a). We only can conclude that a heavy quark may lose more energy when the quark moves perpendicular to the magnetic field. From Fig.~\ref{fig22}(b), we can observe that drag force with a large velocity has a nonmonotonic dependence on temperature and shows an enhancement near the first-order phase transition. It is also found that the chemical potential enhances the energy loss. Moreover, drag force with a large velocity in the perpendicular case is larger than that in the parallel case at the same chemical potential, which is consistent with the results of Fig.~\ref{fig22}(a). The peak value is around $1.3T_c$ and moves toward a lower temperature with increasing chemical potential which is consistent with the phase transition temperature decrease with increasing $\mu$.

\begin{figure}[H]
    \centering
      \setlength{\abovecaptionskip}{0.1cm}
    \includegraphics[width=14cm]{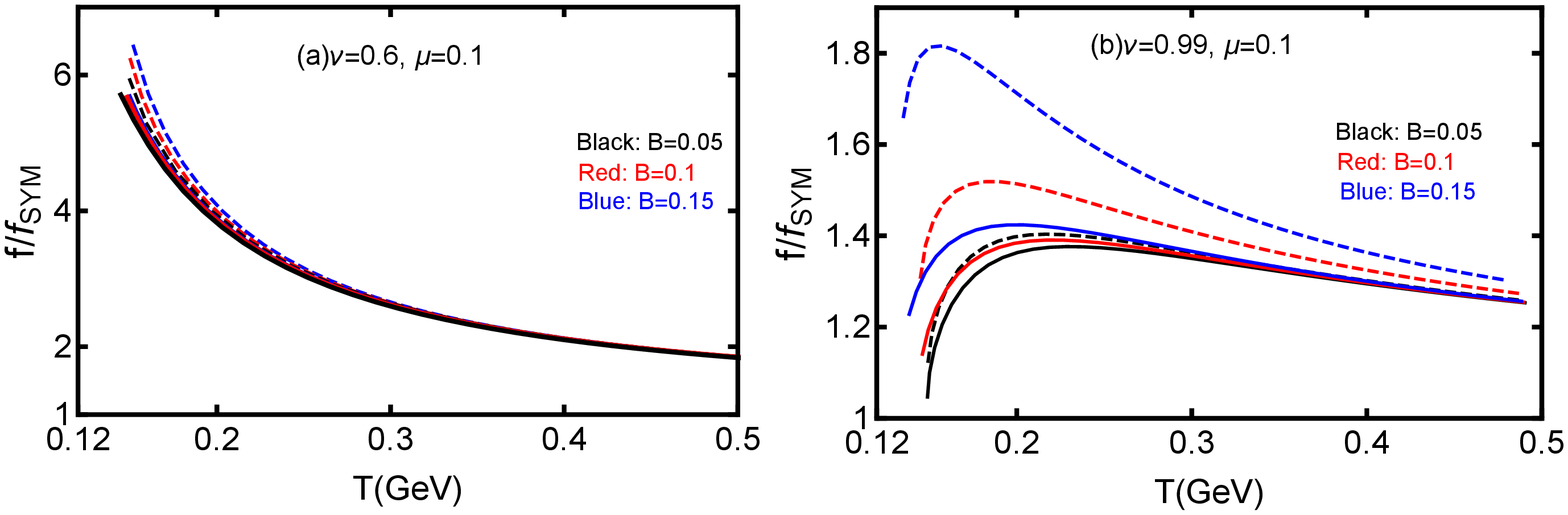}
    \caption{\label{fig23} Drag force normalized by its conformal limit $f/f_{\scriptscriptstyle SYM}$ versus temperature $T$ with different magnetic fields. The solid (dashed) line represents the heavy quark moving parallel (perpendicular) to the magnetic field. The black, red and blue lines denote $B = 0.05, 0.1, 0.15GeV$, respectively. $B$ and $\mu$ are in units GeV.}
\end{figure}

In Fig.~\ref{fig23}, we plot drag force normalized by its conformal limit $f/f_{\scriptscriptstyle SYM}$ versus temperature $T$ with different magnetic fields. From the results, we can find that drag force with small velocity is not sensitive to the magnetic field in Fig.~\ref{fig23}(a). We can observe that the heavy quark loses more energy in the perpendicular case than in the parallel case. From Fig.~\ref{fig23}(b), it is obvious that drag force with a large velocity has a nonmonotonic behavior as a function of temperature and is enhanced near the phase transition temperature. The peak value is moving toward lower temperature with increasing magnetic field which agrees with the phase transition temperature decrease with increasing $B$. The magnetic field enhances heavy quark energy loss. In the perpendicular case, the heavy quark may lose more energy than that in the parallel case at the same magnetic field, which is consistent with the results of Fig.~\ref{fig23}(a).

From the discussion of Figs.~\ref{fig22} and~\ref{fig23}, we find that the heavy quark energy loss is larger in the perpendicular case than in the parallel case. Namely, $f_\perp > f_\parallel$,  which is consistent with the results of the jet quenching parameter $\hat{q}_{\perp}> \hat{q}_{\parallel}$. Jets and heavy quarks have charge and will experience Lorentz force when moving in the magnetized background. The Lorentz force reaches a maximum value when jets or heavy quarks move perpendicularly to the magnetic field. Thus, jets and heavy quarks will lose more energy when moving perpendicularly to $B$.

\section{Conclusion and discussion}\label{sec:06}

In this paper, we consider the holographic QCD model with  nonzero magnetic field and chemical potential. It is found that the metric solutions are self-consistent. Then we discuss the black hole thermodynamics and study the phase diagrams in the $T-\mu$ and $T-B$ planes. We find IMC which is consistent with lattice QCD results. We also discuss the influence of the magnetic field and chemical potential on the location of the CEP. It is found that the magnetic field increases the critical $\mu_{\scriptscriptstyle CEP}$ of the CEP in the  $T-\mu$ plane and the chemical potential increases the critical $B_{\scriptscriptstyle CEP}$ of the CEP in the $T-B$ plane. It is also found that the chemical potential promotes the crossover phase transition into first-order, while the magnetic field promotes the first-order phase transition into crossover.

Further, we find the EOS near the phase transition temperature are nonmonotonic and nontrivial. Then we study the effect of magnetic field on the free energy of the $Q\bar{Q}$ pair with nonzero chemical potential. It is found that the magnetic field suppresses free energy and has a stronger influence when the $Q\bar{Q}$ pair is parallel to the magnetic field. This finding is consistent with the lattice QCD results.

From the analyses of the jet quenching parameter, we find that $\hat{q}/ T^3$ is temperature dependent and has a peak around $1.3T_c -1.4T_c$. Indeed, the peak value of $\hat{q}/ T^3$ moves toward lower temperatures when increasing the magnetic field or chemical potential. This phenomenon is also consistent with the phase transition temperature decrease with increasing $B$ or $\mu$. Moreover, the magnetic field and chemical potential enhance the jet quenching parameter. The chemical potential and magnetic field enhance the energy loss of heavy quarks. Drag force with large velocity is sensitive to the chemical potential and magnetic field and has an enhancement near the phase transition temperature. The peak value moves towards lower temperature with increasing magnetic field which agrees the phase transition temperature decrease with increasing $B$. We also find that the heavy quark energy loss is larger in the perpendicular case than along the magnetic field direction. Namely, $f_\perp > f_\parallel$ which is consistent with the results of the jet quenching parameter $\hat{q}_{\perp}> \hat{q}_{\parallel}$.

We hope our results of the phase diagram and jet quenching parameter in this holographic QCD model can provide more insight into the investigation of phase structure and jet energy loss in heavy ion collision experiments. It is also significant to study the phase transition in the rotating background. We leave this part for future study.

\section*{Acknowledgments}

The authors would like to acknowledge Yan-Qing Zhao for helpful discussion. This work is supported in part by the National Key Research and Development Program of China under Contract No. 2022YFA1604900. This work is also partly supported by the National Natural Science Foundation of China (NSFC) under Grants No. 12435009, and No. 12275104.

\section*{Appendix}

In the appendix, we solve the Einstein-Maxwell equations to justify the metric $Ans\ddot{a}tze$ for small magnetic field, treating the $Ans\ddot{a}tze$ as a small perturbation to the known holographic solutions without magnetic field.

The metric (\ref{eqb}) \cite{Bohra:2019ebj} is
\begin{equation}
\label{eqdd4}
\ ds^{2}=\frac{L^2 S(z)}{z^2}[-g(z)dt^2+dx_{1}^{2}+e^{B^2 z^2}(dx_{2}^{2}+dx_{3}^{2})+\frac{dz^{2}}{g(z)}].
\end{equation}

The factor $e^{B^2 z^2}$ is a natural choice in one of the metric components. The reasons are as follows:
\begin{enumerate}
\item We can find that the energy-momentum tensor is of order $B^2$ from the field strength tensor. Thus, one needs to take the $Ans\ddot{a}tze$ of metric with the $B^2$ term in it.

\item The rotation symmetry $SO(3)$ is recovered when $B$ vanishes.

\item The metric could reduce to AdS at the asymptotic boundary even with finite $B$.

\item It is necessary to mention that $z^2$ term is used to make the exponent dimensionless.

\item The effect of magnetic field on the string tension of $Q\bar{Q}$ in this bottom-up holographic QCD model with the factor $e^{B^2 z^2}$ are more compatible with magnetize lattice QCD results\cite{Bohra:2019ebj}.

\item We find that the magnetic field suppresses the free energy and this suppression is stronger when the connecting line of $Q\bar{Q}$ pair is parallel to the magnetic field than the perpendicular case which is consistent with the lattice QCD results in this work. One also can take the different form of the factor, see \cite{DHoker:2009ixq}.
\end{enumerate}

Then we consider the perturbative expansion in the presence of a small magnetic field
\begin{equation}
\label{eqdd41}
\ ds^{2}\simeq\frac{L^2 S(z)}{z^2}[g_{tt}dt^2+dx_{1}^{2}+(1+B^2 z^2)(dx_{2}^{2}+dx_{3}^{2})+ g_{zz} dz^{2}],
\end{equation}
where
\begin{equation}
\label{eqdd42}
\begin{split}
& g_{tt} = -\overline{g}(z) + B^2 h_{tt}(z),\\
& g_{zz} = \frac{1}{\overline{g}(z)} + B^2 h_{zz}(z),
 \end{split}
\end{equation}
where $\overline{g}(z)$ denotes the blackening function $g(z)$ at $B \rightarrow 0$. $h_{tt}(z)$ and $h_{zz}(z)$ denote the perturbation functions of $g_{tt}$ and $g_{zz}$ respectively, which describe the correction to the metric component.

The perturbative Einstein equation for the $g_{tt}$ component is
\begin{equation}
\label{eqdd43}
\begin{split}
R_{tt}^{(1)} = -\frac{1}{2} \overline{g}(z) h_{tt}''(z) - \frac{3}{2} \frac{\overline{g}(z)}{h(z)} h_{tt}'(z),
 \end{split}
\end{equation}
where $R_{tt}^{(1)}$ is the first-order perturbation of the $tt$ component of the Riccci tensor of the spacetime.

The first-order perturbation of the $zz$ component of the Riccci tensor is
\begin{equation}
\label{eqdd44}
\begin{split}
R_{zz}^{(1)} = \frac{1}{2} \frac{h_{zz}''(z)}{\overline{g}(z)} - \frac{3}{z} \frac{h_{zz}'(z)}{h(z)},
 \end{split}
\end{equation}
where
\begin{equation}
\label{eqdd45}
h(z) = \frac{L^2 S(z)}{z^2}.
\end{equation}

For the electromagnetic field's stress-energy tensor
\begin{equation}
\label{eqdd46}
\begin{split}
& T_{tt}^{(\text{EM})} = \frac{f_2(\phi)}{2} B^2 g^{(0)}_{tt},\\
& T_{zz}^{(\text{EM})} = \frac{f_2(\phi)}{2} B^2 g^{(0)}_{zz},
 \end{split}
\end{equation}
where $g^{(0)}_{tt}$ and $g^{(0)}_{zz}$ are the original $tt$ and $zz$ component of the spacetime metric in the absence of the magnetic field.

The final perturbative equations are
\begin{equation}
\label{eqdd47}
\begin{split}
& -\frac{1}{2} \overline{g}(z) h_{tt}''(z) - \frac{3}{2} \frac{\overline{g}(z)}{h(z)} h_{tt}'(z) = B^2 f_2(z) \overline{g}(z),\\
& \frac{1}{2} \frac{h_{zz}''(z)}{\overline{g}(z)} - \frac{3}{z} \frac{h_{zz}'(z)}{h(z)} = B^2 f_2(z) \frac{1}{\overline{g}(z)},
 \end{split}
\end{equation}
where
\begin{equation}
\label{eqdd48}
f_2(z) = -\frac{L^2 e^{2B^2 z^2 + 2P(z)}}{z} \left[ \overline{g}(z) \left( 4B^2 z + 6P'(z) - \frac{4}{z} \right) + 2\overline{g}'(z) \right],
\end{equation}
where $P(z)= -a\log(b z^2 +1)$.

To determine the integration constants, we specify the boundary conditions as
\begin{equation}
\label{eqdd49}
\begin{split}
& h_{tt}(0) = 0,\  h_{tt}'(0) = 0,\\
& h_{zz}(0) = 0,\ h_{zz}'(0) = 0.
 \end{split}
\end{equation}

With these boundary conditions, the integration constants are zero.

\begin{figure}[H]
    \centering
      \setlength{\abovecaptionskip}{0.1cm}
    \includegraphics[width=9cm]{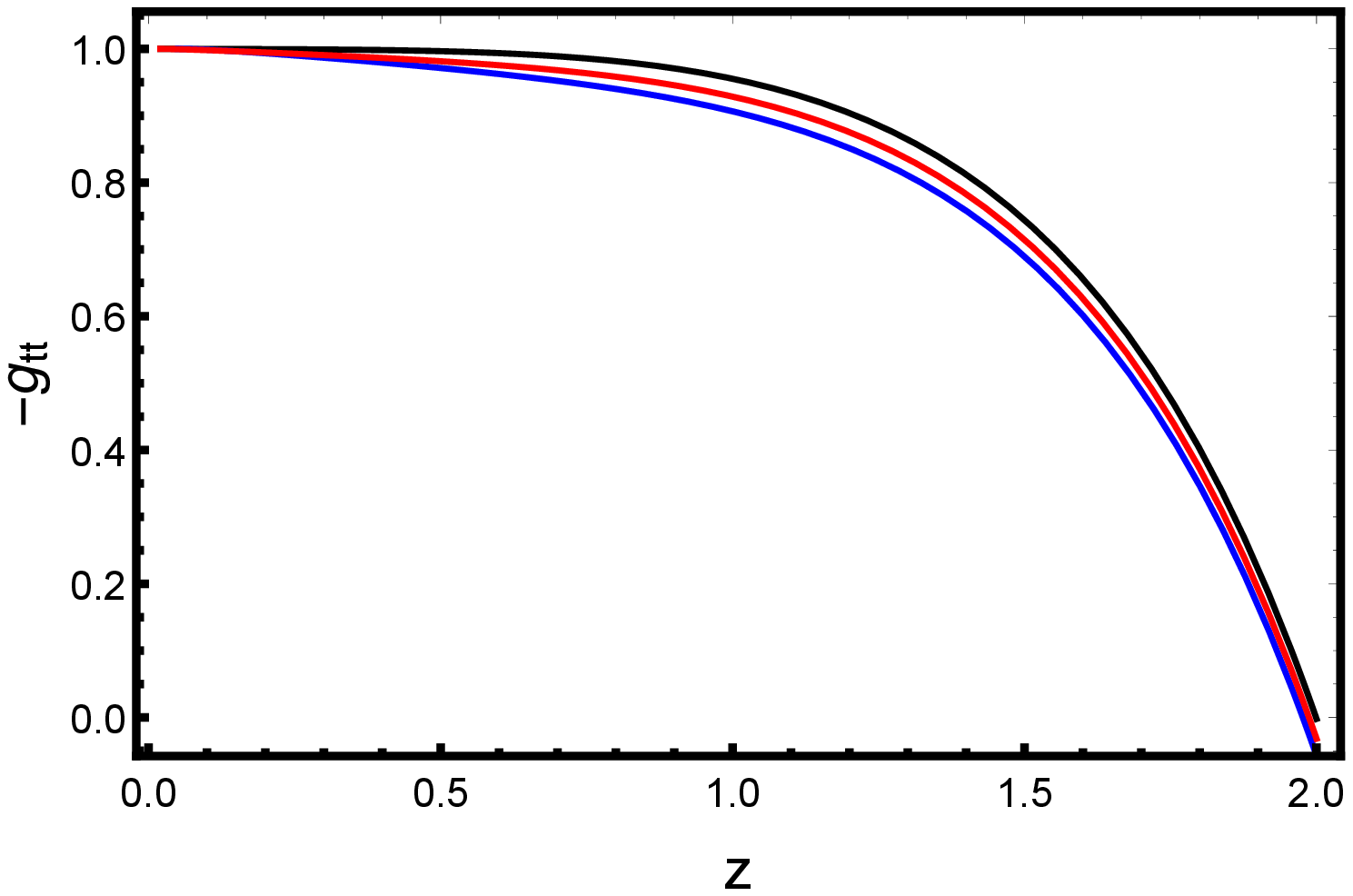}
    \caption{\label{fig24} The effect of magnetic field on $g_{tt}$. The black, red and blue lines denote $B = 0.05, 0.1, 0.15GeV$ respectively.}
\end{figure}

\begin{figure}[H]
    \centering
      \setlength{\abovecaptionskip}{0.1cm}
    \includegraphics[width=14cm]{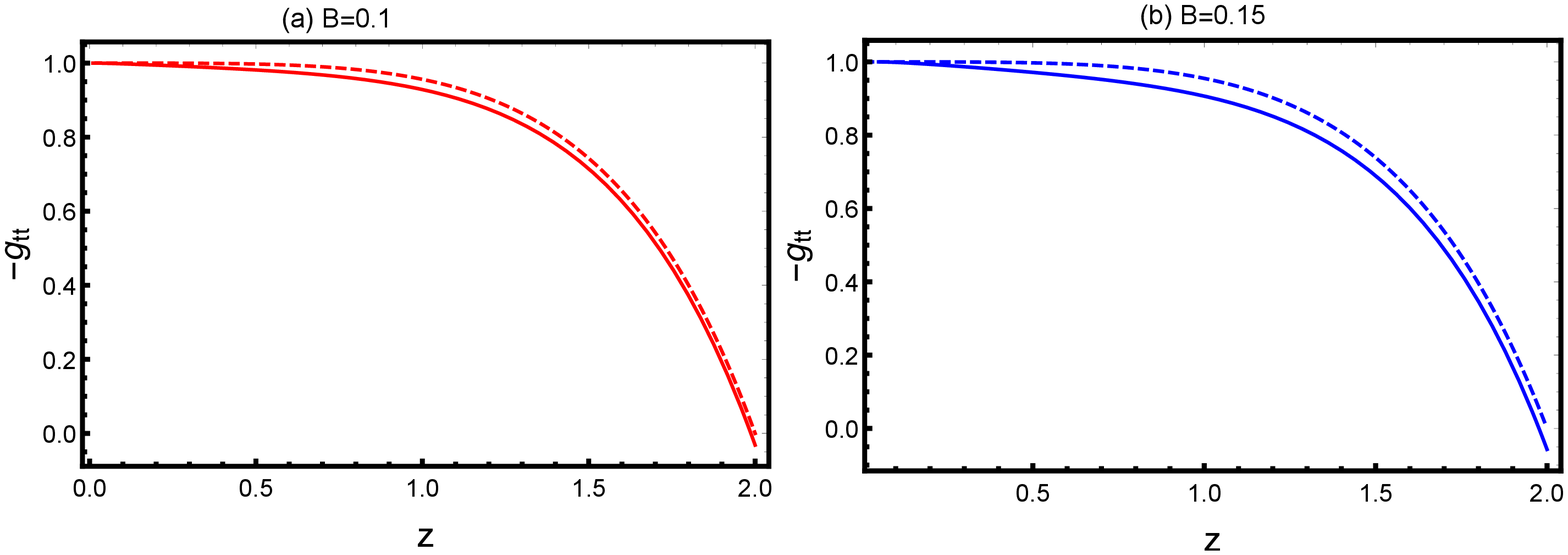}
    \caption{\label{fig25} The difference between the perturbative results (Eq.(\ref{eqdd42})) of $g_{tt}$ and the origin results in the $Ans\ddot{a}tze$ (Eq.(\ref{eqb})) of $g_{tt}$ when B=0.1 and 0.15. The solid line represents the perturbative results, while the dashed line represents the origin results.}
\end{figure}

In order to discuss the influence of $B$, we take $\mu=0$ and $z_h =2$ in the calculations. In Fig.~\ref{fig24}, we plot the effect of magnetic field on the perturbative expansion results (Eq.(\ref{eqdd42})) of $g_{tt}$. One can find that the magnetic field has little influence on the $g_{tt}$. Furthermore, we discuss the difference between the perturbative results (Eq.(\ref{eqdd42})) of $g_{tt}$ and the origin results in the $Ans\ddot{a}tze$ (Eq.(\ref{eqb})) of $g_{tt}$. We can observe that the difference is small.

\begin{figure}[H]
    \centering
      \setlength{\abovecaptionskip}{0.1cm}
    \includegraphics[width=9cm]{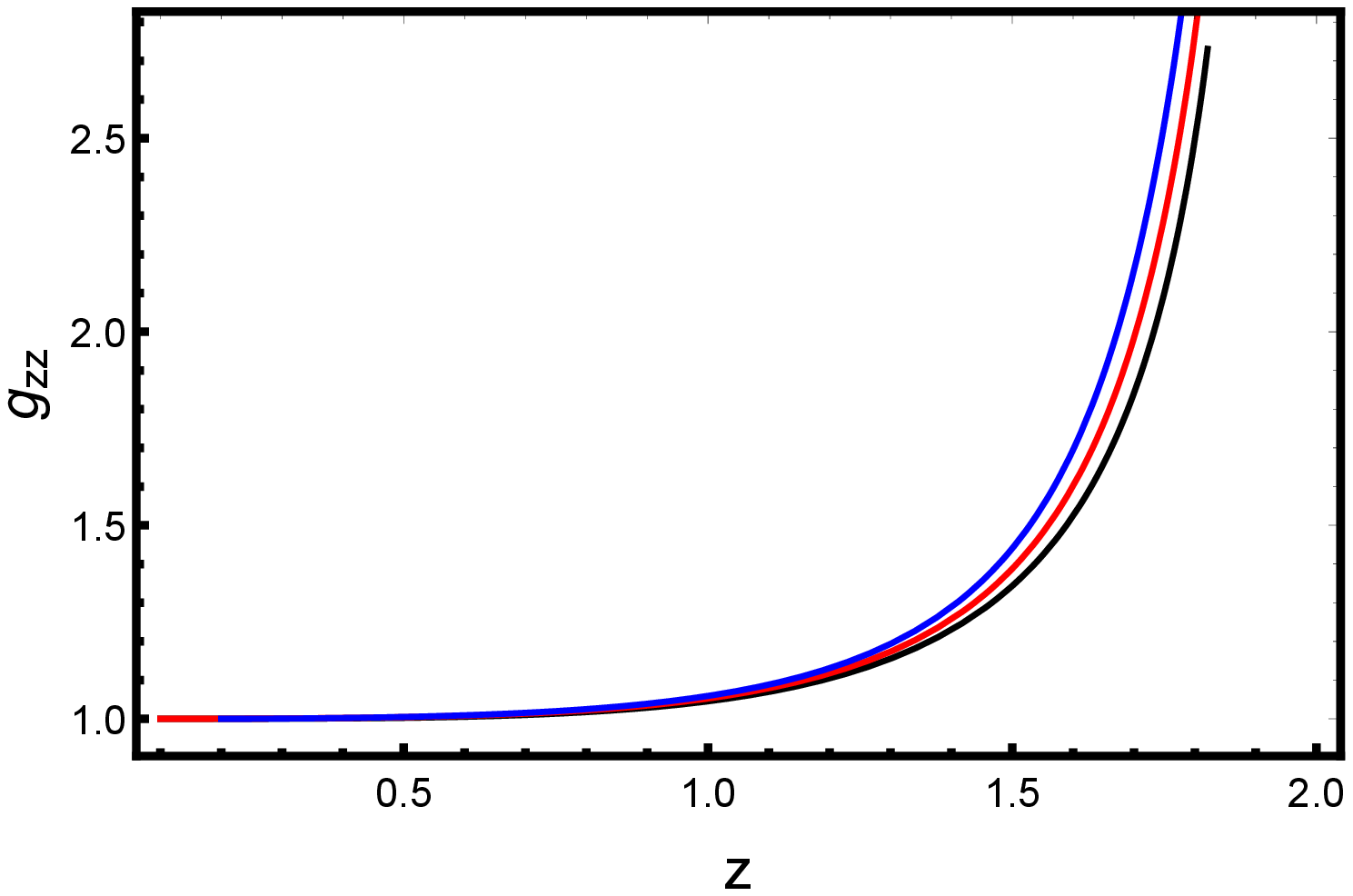}
    \caption{\label{fig26} The effect of magnetic field on $g_{zz}$. The black, red and blue lines denote $B = 0.05, 0.1, 0.15GeV$ respectively.}
\end{figure}

\begin{figure}[H]
    \centering
      \setlength{\abovecaptionskip}{0.1cm}
    \includegraphics[width=14cm]{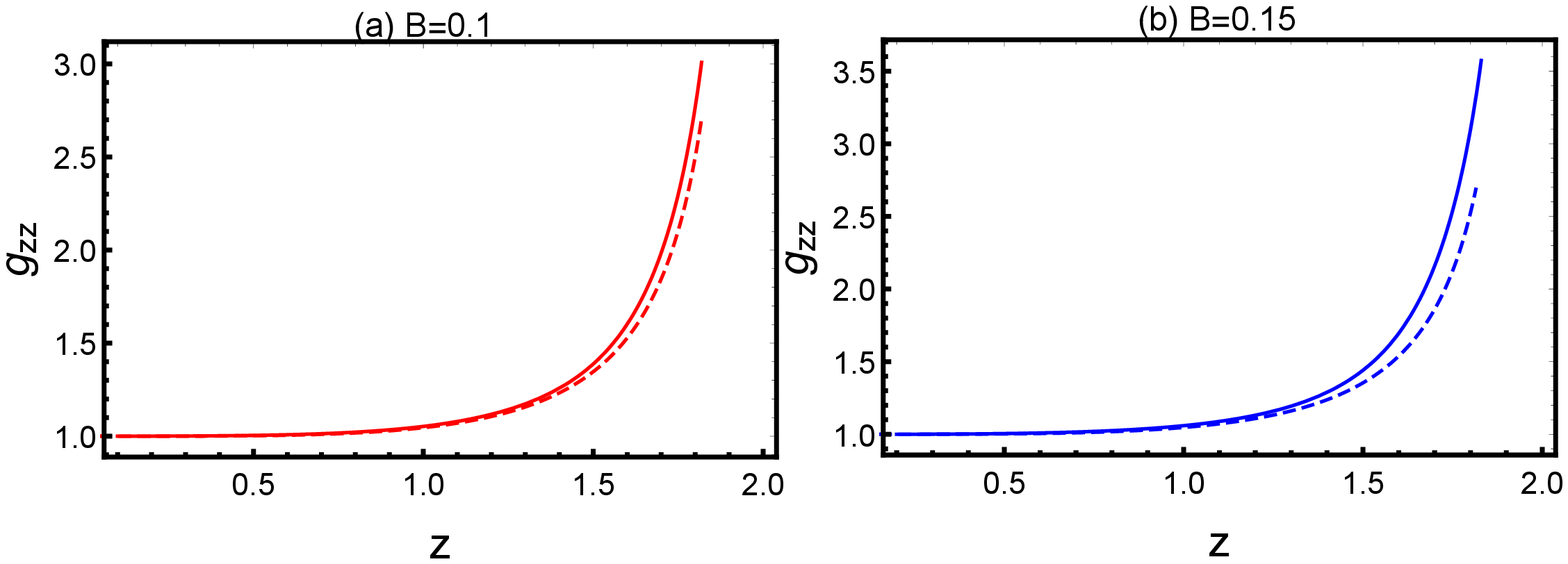}
    \caption{\label{fig27} The difference between the perturbative results (Eq.(\ref{eqdd42})) of $g_{zz}$ and the origin results in the $Ans\ddot{a}tze$ (Eq.(\ref{eqb})) of $g_{zz}$ at B=0.1 and 0.15. The solid line represents the perturbative results, while the dashed line represents the origin results.}
\end{figure}

In Fig.26, we draw the effect of magnetic field on the perturbative expansion results (Eq.(\ref{eqdd42})) of $g_{zz}$. It is found that the magnetic field has little effect on the $g_{zz}$. Moreover, we plot the difference between the perturbative results (Eq.(46)) of $g_{zz}$ and the origin results in the $Ans\ddot{a}tze$ (Eq.(\ref{eqb})) of $g_{zz}$ in Fig.27. We can find that the difference is small.

Through detailed derivation and numerical solution, we verify that under the small magnetic field approximation, the perturbative terms $ h_{tt}(z)$ and $h_{zz}(z)$ conform to our $Ans\ddot{a}tze$ and satisfy the Einstein equations. In Fig.~\ref{fig24}- ~\ref{fig27}, we justify the $Ans\ddot{a}tze$ for the metric for small magnetic field. It is found that the magnetic field has little effect on the perturbative expansion results. Moreover, the difference between perturbative expansion results and the metric (\ref{eqb}) is small. Then one can treat the $Ans\ddot{a}tze$ as a small perturbation to the known holographic solutions.

\end{document}